\documentclass[aps,pre,twocolumn,showpacs]{revtex4-1} 
\usepackage{epsfig}
\usepackage{psfrag}
\usepackage{amsmath}
\usepackage{amsfonts}
\usepackage{amssymb}

\usepackage{graphicx}
\usepackage{caption}

\usepackage{color}
\PassOptionsToPackage{hyphens}{url}\usepackage{hyperref}
\usepackage{breakurl}

\definecolor{RED}{rgb}{1,0,0}
\definecolor{BLUE}{rgb}{0,0,1}

\newcommand{\vc}[1]{\boldsymbol{#1}}
\newcommand{\pack}[0]{\mathcal{P}}
\newcommand{\reals}[0]{\mathbb{R}}

\providecommand \href [0]{\begingroup \@sanitize@url \@href}%

\begin{document}
\title{Maximally dense packings of two-dimensional convex and concave
  noncircular particles}
\author{Steven Atkinson}
\affiliation{Department of Mechanical and Aerospace Engineering, Princeton University, Princeton, New Jersey 08544, USA}
\author{Yang Jiao}
\affiliation{Princeton Institute for the Science and Technology of Materials, Princeton University, Princeton, New Jersey 08544, USA}
\author{Salvatore Torquato}
\affiliation{Department of Chemistry, Department of Physics, Princeton Center for Theoretical Science, Program of Applied and Computational
  Mathematics, Princeton Institute for the Science and Technology of Materials, Princeton University, Princeton New Jersey 08544, USA}
\date{\today}
\begin{abstract}
Dense packings of hard particles have important applications in many
fields, including condensed matter physics, discrete geometry and cell
biology. In this paper, we employ a stochastic search implementation
of the Torquato-Jiao Adaptive-Shrinking-Cell (ASC) optimization scheme
[\href{http://dx.doi.org/10.1038/nature08239} {Nature (London) {\bf 460}, 876 (2009)}] to find maximally dense particle packings in
$d$-dimensional Euclidean space $\mathbb{R}^d$.  While the original
implementation was designed to study spheres and convex polyhedra in
$d \ge 3$, our implementation focuses on $d=2$ and extends the
algorithm to include both concave polygons and certain complex convex
or concave non-polygonal particle shapes.  We verify the robustness of
this packing protocol by successfully reproducing the known putative
optimal packings of congruent copies of regular pentagons and
octagons, then employ it to suggest dense packing arrangements of
congruent copies of certain families of concave crosses, convex and
concave curved triangles (incorporating shapes resembling the
Mercedes-Benz logo), and ``moon-like'' shapes. Analytical
constructions are determined subsequently to obtain the densest known
packings of these particle shapes. For the examples considered, we
find that the densest packings of both convex and concave particles
with central symmetry are achieved by their corresponding optimal
Bravais lattice packings; for particles lacking central symmetry, the
densest packings obtained are non-lattice periodic packings, which are
consistent with recently-proposed general organizing principles for
hard particles.  Moreover, we find that the densest known packings of
certain curved triangles are periodic with a four-particle basis, and
we find that the densest known periodic packings of certain moon-like shapes
possess no inherent symmetries.  Our work adds to the growing evidence
that particle shape can be used as a tuning parameter to achieve a
diversity of packing structures.
\end{abstract}

\maketitle

\section{Introduction}
The problem of packing nonoverlapping particles in $d$-dimensional
Euclidean space $\reals^d$ has been of interest in discrete
mathematics and geometry for centuries.  One overarching aim is to
ascertain organizing principles that govern the nature of dense
packings of various shapes \cite{OrganizingPrinciples} in order to
better understand many natural phenomena, including liquid, glassy and
crystalline states of matter
\cite{BernalLiquids,ZallenAmorphous,HolyBible,CondensedMatterPhys};
heterogeneous materials \cite{HolyBible}; crystalline polymers
\cite{PackPolymers,Polymers2}; and biological systems
\cite{PackProteins,Viruses,BrainTissue}; to name a few.  In two
dimensions, the packing of hard particles has implications for
understanding the behavior and structures found in thin films
\cite{ThinFilmsBook}, adsorption of molecules on substrates
\cite{PackingSubstrateSTM,SelfAssemblyPackingSubstrate}, and the
organization of epithelial cells
\cite{Farhadifar20072095,PackingFlyWing}. 

In the two-dimensional Euclidean plane $\reals^2$, considerable effort
has been devoted to study and characterize packings (roughly speaking,
large collections of nonoverlapping particles), especially of
congruent copies of convex particles
\cite{RogersBook,RogersLattice,FT1950,TilingsAbound,ConwayPolyominoes,ConwayQuasiTile}.
Perhaps the simplest characteristic of a packing is its packing
density, $\phi$, which is, intuitively speaking, the fraction of the
plane covered by the particles.  It is well known that the triangular 
lattice is the densest packing of congruent circles, and its packing
density is $\phi = \pi / \sqrt{12} = 0.906899 \dots$
\cite{FTCirclePacking}.  Other simple two-dimensional shapes that have
been studied include the class of regular polygons and related
variations.  For example, it is known that the densest packing of
congruent regular octagons is the optimal Bravais lattice packing with
$\phi = \left[4 (3 -  \sqrt{2} )\right] / 7 = 0.906163 \dots$.  When
the corners of the octagon are all appropriately ``smoothed'' into
hyperbolic curves, the resulting ``smoothed octagon'' is conjectured
to possess the lowest optimal packing density among all convex,
centrally-symmetric particle shapes, with $\phi = (8 - 4 \sqrt{2} -
\ln 2) / (2 \sqrt{2} - 1) = 0.902414 \dots$
\cite{ReinhardtSmoothedOctagon}.  Another simple yet interesting case is 
the regular pentagon, which has a putative maximum packing density of $\phi =
(5 - \sqrt{5} ) / 3 = 0.921311 \dots$, given by its densest
(non-Bravais) double-lattice packing
\cite{HenleyPentagons,KuperbergDoubleLattice,PentagonExperiment,FrenkelPentagons}.

Generally, it has been shown independently by both Fejes T\'oth
\cite{FT1950} and Rogers \cite{RogersLattice} that the densest packing
of congruent copies of any convex, two-dimensional shape possessing
central symmetry is achieved by a lattice structure.  Fejes T\'oth
\cite{FTBound} and Mahler \cite{Mahler} also proved that such a
construction must be able to achieve a packing density of at least
$\phi = \sqrt{3} / 2 = 0.866025 \dots$ for any convex shape.
Moreover, Kuperberg and Kuperberg showed that, for any convex shape
(with or without central symmetry), a double-lattice packing may be
constructed, also with $\phi \ge \sqrt{3} / 2$, and conjectured that
the densest double-lattice packing realizes the maximum packing
density for shapes such as regular pentagons and heptagons while
asking if this might extend to all regular polygons with an odd number
of sides \cite{KuperbergDoubleLattice}.

In the present work, we use a two-dimensional stochastic search
implementation of the Torquato-Jiao Adaptive Shrinking Cell (ASC)
packing method, originally implemented in three and higher dimensions
in Refs.\ \cite{YangNature,YangPlatoArchie,TJ}, to generate the
densest known packings of a variety of nontrivial convex and concave
noncircular particles.  The ASC scheme generates dense packings by
rearranging a nonoverlapping configuration of particles within a
periodic Fundamental Cell (FC) whilst decreasing its volume to
increase the packing density.  Our current implementation of the ASC
scheme expands upon the previous one by detecting overlaps between any
concave polygons and certain smoothly shaped convex and concave
nonspherical particles. This allows us to readily investigate dense
packings of a wide spectrum of nontrivial nonspherical particles in
the plane that exhibit unique packing behaviors.

In this paper, we study the dense packing behavior of several families
of noncircular particle shapes including so-called ``fat crosses'',
convex and concave ``curved triangles'' (incorporating shapes
resembling the Mercedes-Benz logo), and ``moon-like'' shapes.  We then
use the results of the ASC algorithm to inform analytical
constructions of the densest known packings of those particle shapes
considered, recovering several well-known results, then moving onward
to study other particle shapes.  Specifically, we find that the
densest packings of certain curved triangles are periodic with a
four-particle basis.  Also, we find that the densest periodic packings
of certain moon-like shapes possess no inherent symmetries.  Our work
adds to the growing evidence that particle shape can be used as a
tuning parameter to achieve a diversity of packing structures.

The rest of this paper is organized as follows: in
Sec. \ref{sec:Definitions}, we will introduce the mathematical
definitions that are necessary for a rigorous treatment of the packing
problem; in Sec. \ref{sec:Algorithm}, we review the ASC scheme and
discuss our contributions to the method; in Sec. \ref{sec:Shapes}, we
introduce the particle shapes whose packing behavior we have studied;
and in Sec. \ref{sec:Results}, we discuss a number of results in
packing congruent copies of both well-known and other particle shapes.
We then offer conclusions and plans for future work.  In addition, an
appendix includes some additional numerical results, and the
{\it Supplemental Material} contains mathematical details for many of
the packing structures presented in this work\footnote{See Supplemental
  Material at \url{URL here} for mathematical details for the packing
  structures presented in this paper.}.

\section{Definitions}
\label{sec:Definitions}
In order to make precise the problem that will be addressed, we
introduce some mathematical definitions.  First, we define a particle,
$S$, to be a closed, simply-connected set in $\reals^2$ that may be
either concave or convex.  The boundary of the set is denoted as
$\Gamma$.  A special case of convexity is strict convexity, denoting a
convex boundary that contains no line segments.  The area of $S$ is
denoted by $a_1$, and we will henceforth assume that this quantity is
bounded.

\begin{figure}[hbt]
  \begin{centering}
    $\begin{array}{cc}
      \includegraphics[width=0.25\textwidth]{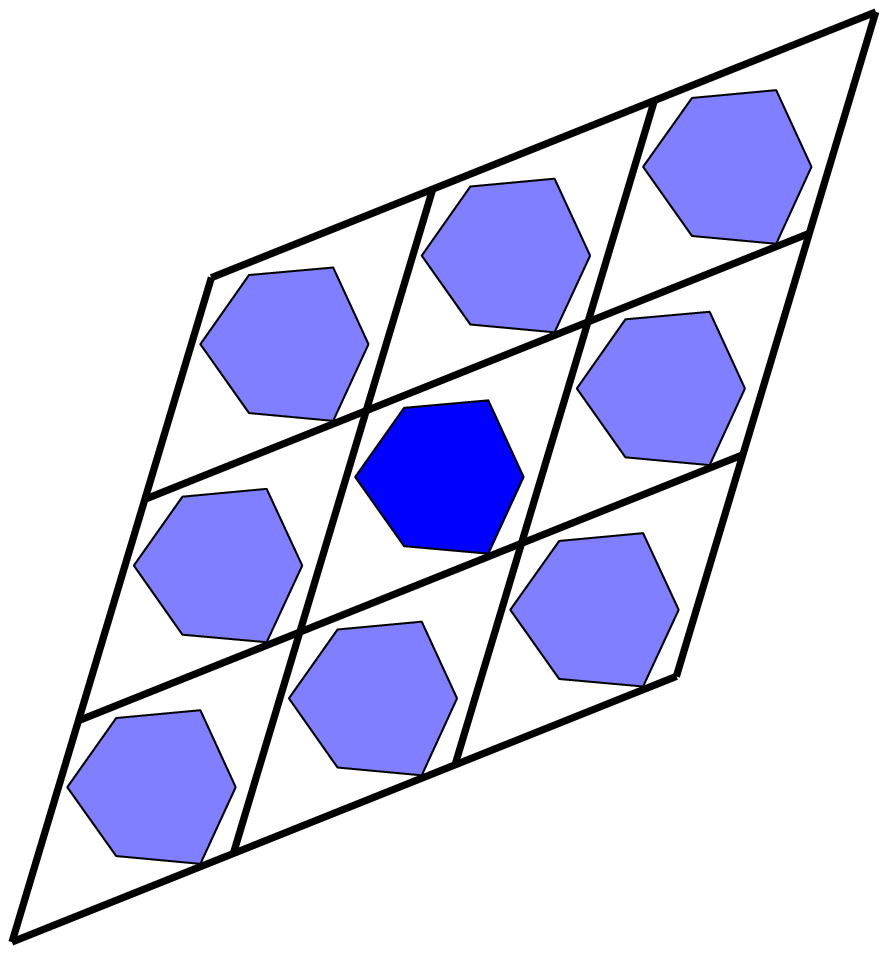}
      &
      \includegraphics[width=0.25\textwidth]{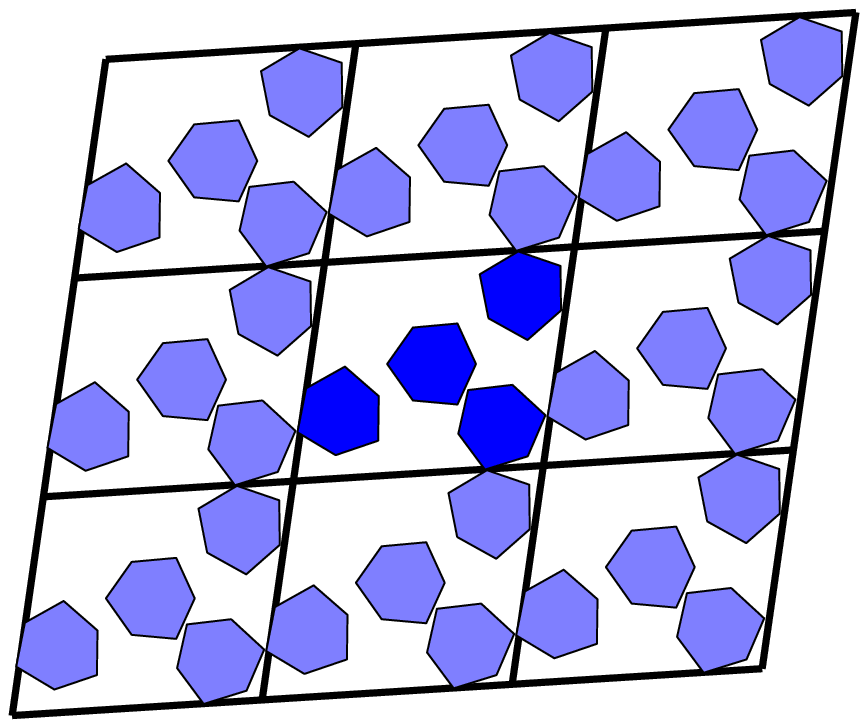}
      \\
      \mbox{(a)}
      &
      \mbox{(b)}
    \end{array}$
    \caption{(Color online) Examples of a lattice packing (a) and periodic 
      packing with four-particle basis (b).  The lattice parallelogram is shown by the black grid
      behind the particles.}
    \label{fig:LatticeVsPeriodic}
  \end{centering}
\end{figure}

Given two linearly independent (column) vectors $\vc{\lambda}_1$ and
$\vc{\lambda}_2$, the {\it lattice} generated by $\vc{\lambda}_1$ and
$\vc{\lambda}_2$ is defined as the set $\left\{i \vc{\lambda}_1 + j
\vc{\lambda}_2~\forall i,j \in \mathbb{Z} \right\}$.  A packing,
$\pack$ is defined as a collection of particles $\{ S_i \}$ whose
interiors are mutually disjoint. If all members of $\pack$ are
translates of each other where the vectors of translation form a
lattice, $\pack$ is known as a {\it Bravais lattice} (or, simply, {\it
  lattice) packing}.  Furthermore, if $\pack$ can be decomposed into
the union of two distinct lattice packings $\pack_0$ and $\pack_1$,
such that an inversion about some point in the plane interchanges
$\pack_0$ and $\pack_1$, then $\pack$ is called a {\it double-lattice
  packing}.  Generally, if one can decompose $\pack$ into the union of
$N \ge 1$ distinct lattice packings, each sharing the same lattice
vectors, then $\pack$ is said to be a {\it periodic packing} with an
{\it $N$-particle basis}.  Also, for all periodic packings, there
exists a fundamental cell (FC) of the packing, parallelogrammatic in
shape, described by a lattice matrix $\mathbf{\Lambda} = \left\{
\vc{\lambda}_1 , \vc{\lambda}_2 \right\}$, inside which all $N$
centroids lie.  Examples of lattice and periodic packings are given in
Figure \ref{fig:LatticeVsPeriodic}.  Note that, in a lattice packing,
all particles must have the same orientation, whereas, in a general
periodic packing, the $N$ particles in the FC are free to have their
own orientations.

The packing density, $\phi$, defined for a given $\pack$ is,
intuitively speaking, the fraction of the plane covered by the copies
of $S$.  When it is assumed that $\pack$ is a periodic packing with an
$N$-particle basis,
\begin{eqnarray}
\phi = \frac{N a_1}{{\rm Area}(F)}
\end{eqnarray}
, where $a_1$ denotes the area of a single particle, and 
${\rm Area}(F)$ denotes the area of the FC.  If $\phi = 1$, then 
$\pack$ is said to be a {\it  tiling}.

\section{Torquato-Jiao Adaptive Shrinking Cell (ASC) Optimization
  Scheme}
\label{sec:Algorithm}
The Torquato-Jiao Adaptive Shrinking Cell (ASC) optimization scheme
seeks to generate dense packings of a collection of shapes within a
periodic FC through a process of rearranging the positions of the
shapes within an FC while decreasing the FC's volume in order to
increase the packing density.  Formally stated, the ASC optimization
scheme is
\begin{eqnarray}
{\rm minimize } &~& -\phi(\vc{r}_1^\lambda , \vc{r}_2^\lambda ,
\vc{r}_3^\lambda , \dots , \vc{r}_N^\lambda ; \theta_1 , \theta_2 ,
\theta_3 , \dots , \theta_N ; \vc{\Lambda}),
\nonumber
\\
{\rm such ~ that} &~& (S_i \cap S_j) \subseteq (\Gamma_i \cup
\Gamma_j) ~ \forall i,j = 1,2,3,\dots,N,
\nonumber
\\
i \not = j,
\end{eqnarray}
where $N$ is the number of particles in the FC, and $\vc{r}^{\lambda}_i$ and $\theta_i$ specify the position and orientation of particle $i$, respectively (see below).  The optimization
scheme can be solved using a variety of techniques including
stochastic search methods with simulated annealing
\cite{YangNature,YangPlatoArchie} and linear programming \cite{TJ};
the present work uses an adaptation of the former.  For the sake of
completeness, the technique will be described here.

In $\reals^2$, the ASC scheme utilizes a parallelogrammatic
fundamental unit cell (FC) with periodic boundary conditions.  The
positions of the $N$ particles are given by the lattice coordinates
(i.e., the relative coordinates with respect to the lattice vectors)
of their centroids, $\vc{r}_1^\lambda , \vc{r}_2^\lambda ,
\vc{r}_3^\lambda , \dots \vc{r}_N^\lambda \in [0,1)^2$; and (global)
  orientations, $\theta_1 , \theta_2 , \theta_3 , \dots , \theta_N \in
  [0,2 \pi)$.  From the initial configuration, the stochastic search
    method uses an iterative process to increase the packing density.
    Its main steps are the following:
\begin{itemize}
\item Random rotations or translations are applied to the shapes,
  accepting moves that satisfy the required nonoverlap constraints
  between the shapes (``random movements''), then
\item Random strains, composed of a combination of a deformation and
  either a dilation or a compression, are applied to the FC that seek
  to either increase or decrease its area with a specified
  probability, corresponding to uphill and downhill moves,
  respectively (``random strains'').
\end{itemize}

During the ``random movements'' step, every particle in the basis is
either translated or rotated within some prescribed limits on the
magnitude of the movement.  If the new position of the particle does
not cause it to overlap with any other particles in the FC (or their
periodic images), the move is kept; otherwise, the move is rejected
and the particle remains where it began.  Note that only one
particle is moved at a time; collections of particles never move
simultaneously during this step.  The process is repeated a specified
number of times for each particle to explore the configurational space
of the packing; the precise number is determined empirically based on
the criterion that the particles are allowed to equilibrate before
attempting to strain the FC.  In the present work, at least 500 trial
movements are attempted for each particle at each occurrence of this
step, and this number remains constant for the duration of the
simulation.

During the ``random strains'' step, the simulation box is
simultaneously deformed and compressed or dilated in a way that
attempts to decrease its area on average while preserving the
nonoverlap constraints.  Since the locations of the particles are
expressed in terms of the lattice vectors, straining the simulation
box also effects a collective motion of the particles.  The straining
process is attempted up to a prescribed maximum number of attempts.
The first successful strain is kept, and the algorithm returns to the
first step (random movements).  After each unsuccessful strain
attempt, the maximum allowed strain is decreased by a constant ratio
in order to steadily increase the chances of finding a valid strain.
Therefore, more attempts are required as the packing increases in
density.  In addition, uphill moves that allow the FC to expand are
allowed with a given probability.

The maximum magnitude of the trial movements is steadily decreased
throughout the execution of the algorithm, reducing the maximum
magnitude by a constant ratio when the acceptance rate of trial moves
falls significantly below 50\%.  Moreover, while the maximum strain
magnitude is decreased after each unsuccessful attempt, the maximum
magnitude is restored to its original value the next time the ``random
strains'' step occurs, since, for example, particle movements within
the FC may result in two particles being next to each other at the end
of one ``random movements'' step (necessitating a small strain), and
may result in the particles being more uniformly spaced at the end of
the next step (allowing for a larger strain even though the FC may be
smaller).

The sequence of random movements and random strains is repeated a
prescribed number of times, chosen such that the algorithm has
``enough time'' to find and settle in a minimum of the objective
function (which is determined by monitoring the convergence of the
packing density); in the present work, the maximum number of
iterations is always at least 500.  In order to refine the results, a
 ``fine tuning'' procedure may be used in which the ASC
algorithm is executed a second time, starting with the previous final
dense configuration by expanding the FC by some small amount and
greatly reducing the magnitude of the movements and strains in order
to more accurately approach the density maximum that has been
identified.  However, this process is only used in cases where the
particles are convex, since this is necessary to ensure that a
general expansion of the FC does not introduce overlaps.

In the past, the ASC scheme has been applied to determining packings
of convex polyhedra in three and higher dimensions, detecting overlaps
between particles through use of the separation axis theorem
\cite{YangNature,YangPlatoArchie}; in this work, the algorithm has
been modified to detect overlaps between concave particles in two
dimensions through a combinatoric check of the particle's edges, which
may be some combination of line segments and circular arc segments.

This solution of the ASC scheme is a particularly strong investigative
tool since it is capable of identifying dense packing configurations
quickly and with appreciable consistency.  Analytical methods may then
be used, benefitting from the information derived from the algorithm's
results to determine the exact structures of the packings.  In
addition, the freedom to choose any initial condition may be used to
enhance the quality of solutions found if an interesting structure is
more often realized through specific initial conditions \cite{TJ}.  In
the present work, we find that, by choosing initial lattice vectors
that more closely resemble the final lattice vectors of dense
structures, denser results are achieved, and with increased
frequency.

\section{Noncircular Particle Shapes}
\label{sec:Shapes}
A variety of noncircular particle shapes were studied in the present
work; this section aims to make precise the geometries that were
studied.  In order to establish benchmarks for the program's
performance, the well-known cases of regular pentagons and octagons
were considered.  In addition, we consider three families of nontrivial particle
shapes: ``fat crosses'', ``curved triangles'', and ``moon-like
shapes''. The packing characteristics of these noncircular particle
shapes have heretofore not been studied. We show that they lead to
unique packing arrangements.

\subsection{Fat Cross}
\begin{figure*}[hbt]
  \begin{centering}
    \psfrag{l}{$l$}
    \psfrag{d}{$\delta$}
    $\begin{array}{ccc}
      \includegraphics[width=1.5in]{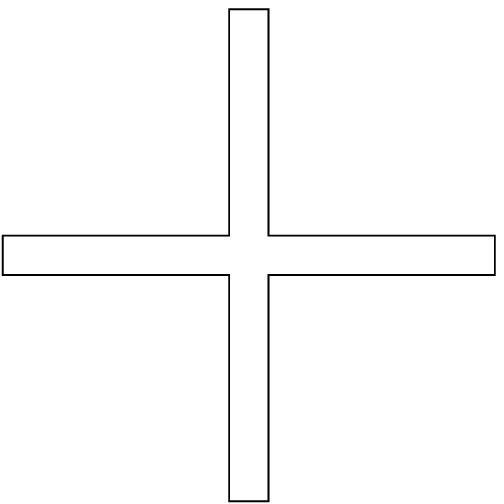}
      &
      \includegraphics[width=1.5in]{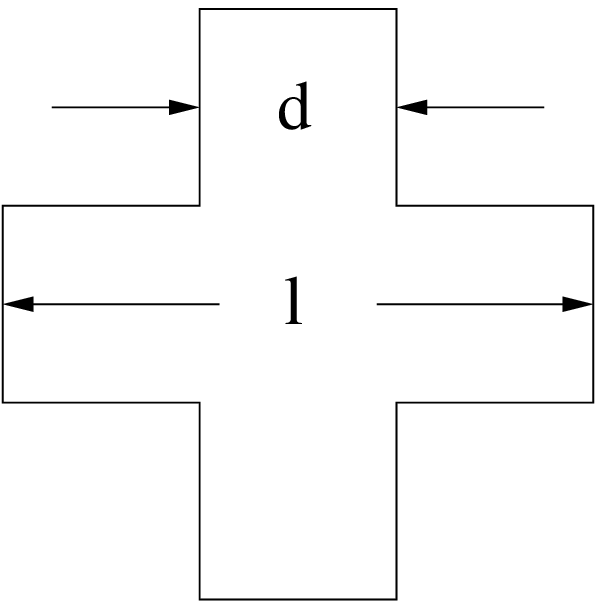}
      &
      \includegraphics[width=1.5in]{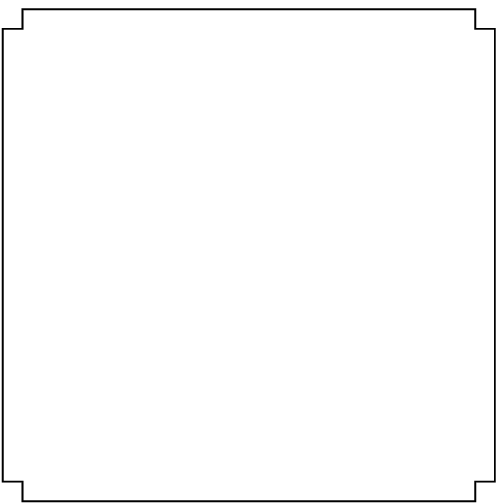}
      \\
      \mbox{(a)}
      &
      \mbox{(b)}
      &
      \mbox{(c)}
    \end{array}$
    \caption{Three instances of the general ``fat cross'' with $w =
      1/10$ (a), $1/3$ (b), and $9/10$ (c).}
    \label{fig:Cross}
  \end{centering}
\end{figure*}

The first nontrivial particle shape that we investigate is a so-called
``fat cross.''  Several examples of this particle shape are given in
Figure \ref{fig:Cross}.  The particle shape is described by a width
parameter $w$ according to the definition 
\begin{eqnarray}
w = \frac{\delta}{l},
\end{eqnarray}
where $\delta$ is the width of the cross's legs, and $l$ is the
end-to-end length of the cross. Therefore, it follows that $w \in
[0,1]$; the $w = 0$ case corresponds to a pair of lines of length $l$
bisecting each other at a right angle, and the $w = 1$ case
corresponds to a square of side length $l$. 

\subsection{Curved Triangle}
\begin{figure*}[hbt]
  \begin{centering}
    \psfrag{r}{$r$}
    \psfrag{r0}{$r_0$}
    $\begin{array}{c@{\hspace{0.5in}}c}
      \includegraphics[width=2.5in]{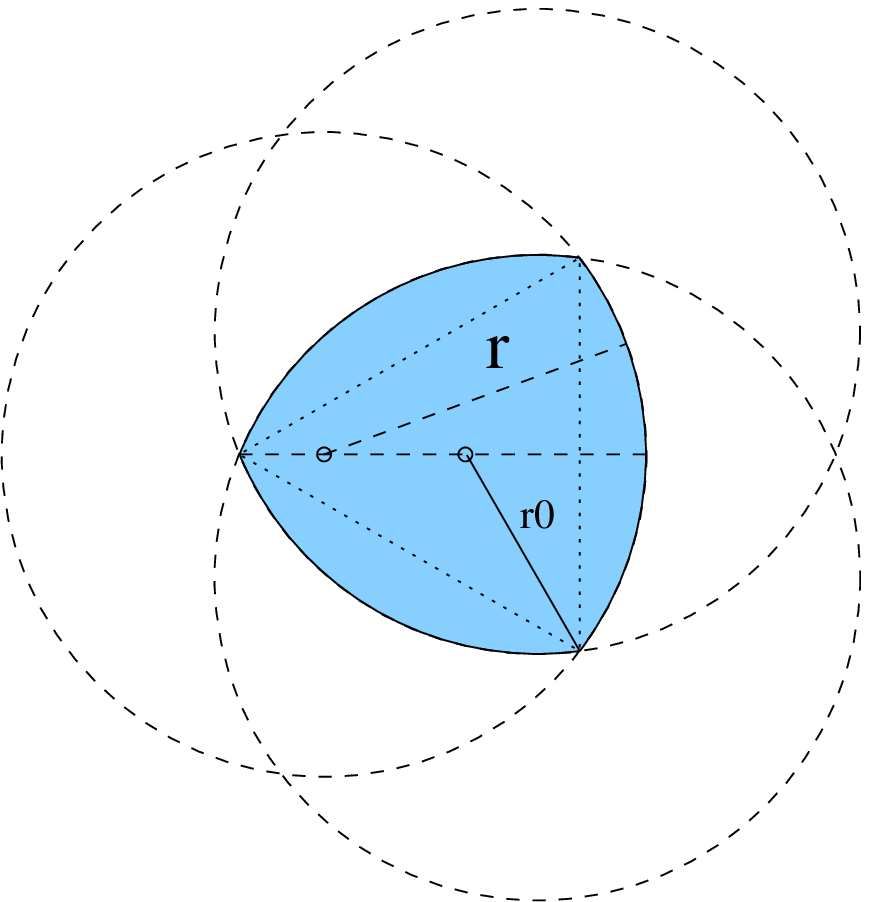}
      &
      \includegraphics[width=2.5in]{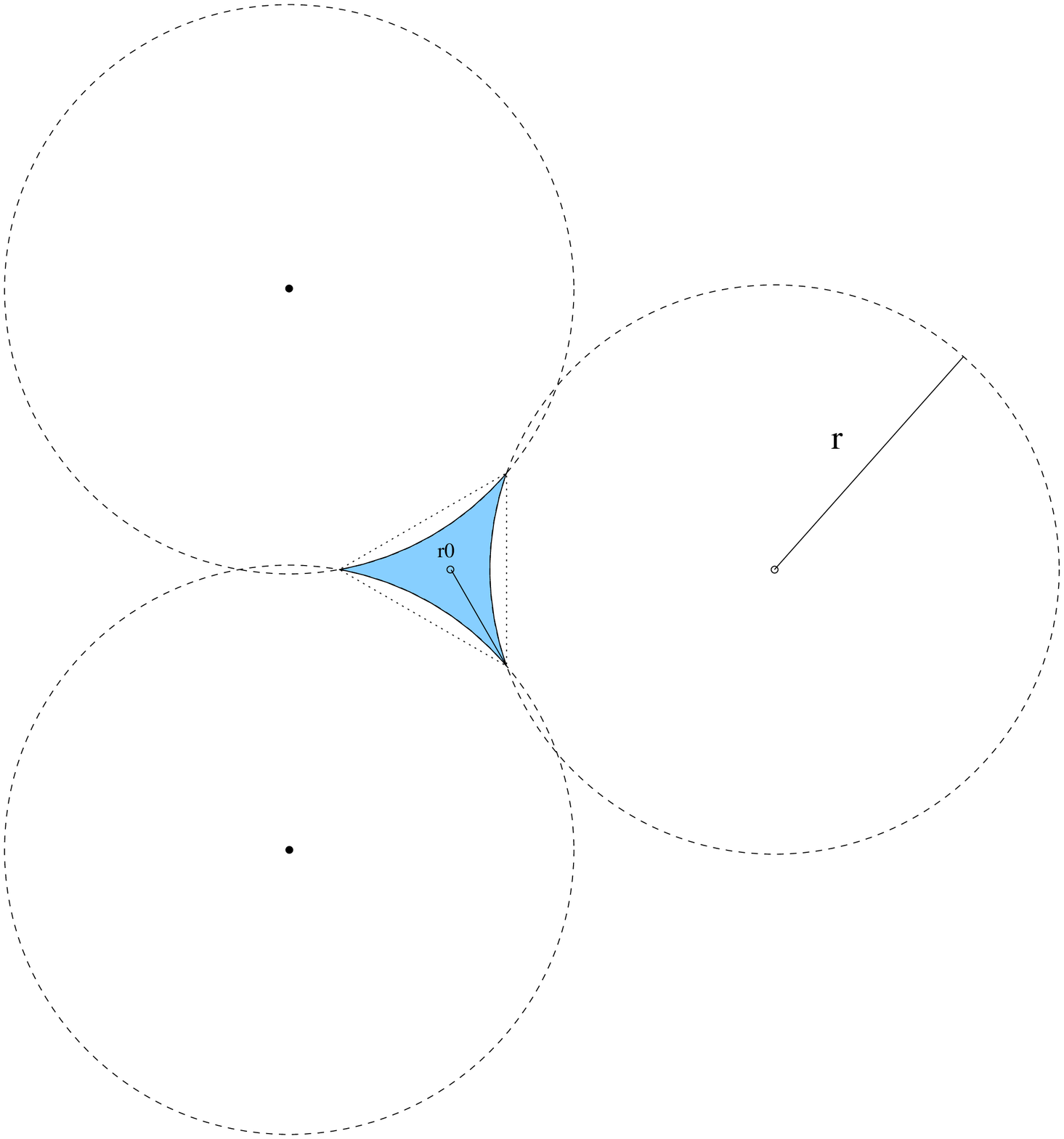}
      \\
      \mbox{(a)}
      &
      \mbox{(b)}
    \end{array}$
    \includegraphics[width=2.5in]{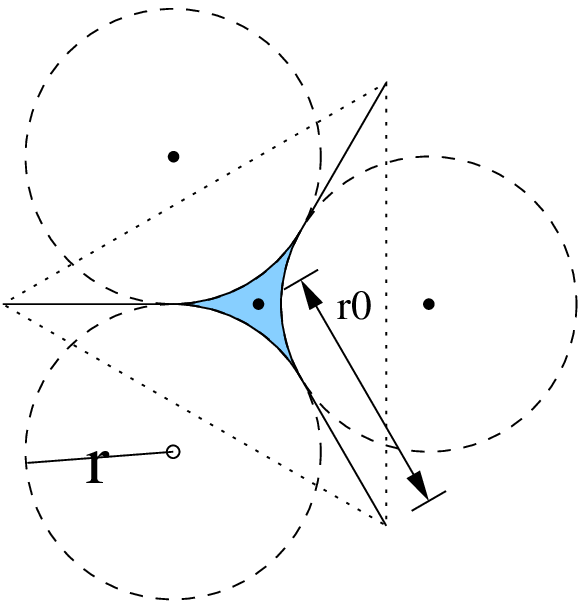}
    \\
    \mbox{(c)}
    \caption{(Color online) Three examples of the basic form of a
      curved triangle: convex ((a), $k =  0.71$), concave ((b), $k =
      -0.39$), and ``spiked'' ((c), $k = -1.55$).}
    \label{fig:CurvedTriangle}
  \end{centering}
\end{figure*}

Another particle shape that we investigate is the so-called ``curved
triangle,'' that, in the convex case, is a two-dimensional analog of
the ``tetrahedral puff'' described in \cite{Yoav}.  This particle
shape is derived by replacing the sides of an equilateral triangle
with circular arc segments; this is illustrated in Figure
\ref{fig:CurvedTriangle}.  The particle shape may then described by a
parameter
\begin{eqnarray}
k = \pm \frac{r_0}{r},
\label{eqn:CT}
\end{eqnarray}
where $r$ is the radius of the arc segments, and $r_0$ is the radius
of a circle passing through the triangle's vertices.  $k$ is taken by
convention to be positive when the curved triangle is convex as in
Figure \ref{fig:CurvedTriangle}a, and negative when the curved triangle
is concave as in Figure \ref{fig:CurvedTriangle}b. When $k = 0$, then
the particle shape becomes an equilateral triangle; when $k = 1 /
\sqrt{3} = 0.577350 \dots$, the particle shape becomes the well-known
Reuleaux triangle \cite{ReuleauxBook}; and when $k = 1$, the particle
shape becomes a circle.  When $k = -1 / \sqrt{3}$, the vertices of the
particle become cusps, and in order to decrease $k$ beyond this point,
a line segment connects the cusp and the vertex, as is shown in Figure
\ref{fig:CurvedTriangle}c; the curved triangle is said to be ``spiked''
in this regime, and begins to resemble the Mercedes-Benz logo.  In the
limit $k \rightarrow - \infty$, the particle shape becomes the union
of three line segments of length $r_0$, each at an angle of $2 \pi /
3$ from the others.

\subsection{Moon-Like Shape}
\begin{figure}[hbt]
  \begin{centering}
    \psfrag{r}{$r$}
    \psfrag{r0}{$r_0$}
    $\begin{array}{c@{\hspace{0.5in}}c}
      \includegraphics[width=0.2\textwidth]{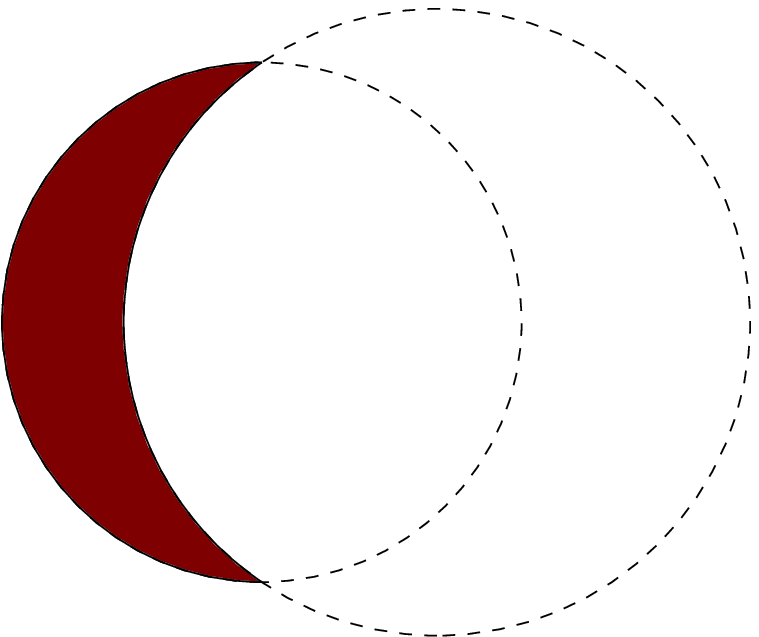}
      &
      \includegraphics[width=0.2\textwidth]{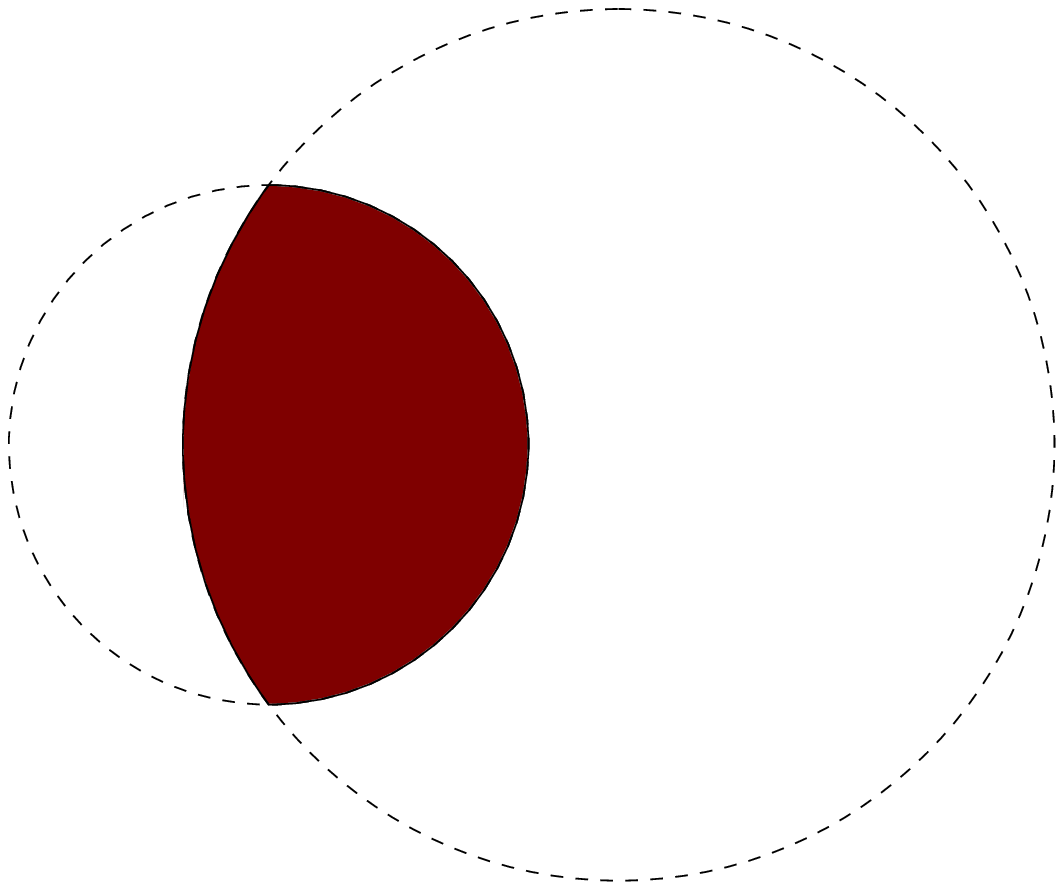}
      \\
      \mbox{(a)}
      &
      \mbox{(b)}
    \end{array}$
    \caption{(Color online) Examples of a ``crescent'' ((a), $k > 0$) and a
      ``gibbous'' ((b), $k < 0$) moon-like shape, .}
    \label{fig:Moon}
  \end{centering}
\end{figure}

The third nontrivial particle shape that we investigate is the a
``moon-like'' shape, which is constructed by a pair of circular arc
segments with radii $r_0$ and $r$, where the former arc is a
half-circle.  The particle shape is parameterized by the same
parameter $k$ used to characterize the curved triangle [cf. Eq. (\ref{eqn:CT})],
where $k$ is, by convention, taken to be negative when the particle
shape is a concave, ``crescent'' shape, as in Figure
\ref{fig:Moon}a and positive when the moon is a convex,
``gibbous'' shape, as in Figure \ref{fig:Moon}b.

\section{Results}
\label{sec:Results}
The stochastic search solution of the ASC scheme described in
Sec. \ref{sec:Algorithm} is used to generate dense periodic packings
of congruent copies of the particle shapes described above, and the
resulting computer-generated packings are used to inform analytical
predictions for the structures of the densest packings; the following
details the results of the numerical simulations and discusses the
structures that we find.  We also provide visual representations of
some noteworthy cases.  A summary of the densest packing behavior of
the noncircular particle shapes the we obtain is given in Table
\ref{tab:Summary}.

\begin{table*}[t]
\begin{center}
\caption{Symmetries of the putative densest packings of the
  aforementioned particle shapes.  The properties of well-known shapes
  (pentagons and octagons) are compared to the results we obtain for
  our nontrivial particle shapes (fat crosses, curved triangles, and
  moon-like shapes).}
\begin{tabular}{lccccccc}
\hline \hline
Shape Name & \hspace{5 mm} & Convex & \hspace{5 mm} & Centrally-Symmetric Particle  &
\hspace{5 mm} & Centrally-Symmetric Basis & $N$ \\ \hline 
Pentagon        && Yes            && No  && Yes            & 2 \\ 
Octagon         && Yes            && Yes && Yes            & 1 \\ 
Fat Cross       && No             && Yes && Yes            & 1 \\ 
Curved Triangle && depends on $k$ && No  && Yes            & 2,4 (depending on $k$) \\ 
Moon-like shape && depends on $k$ && No  && depends on $k$ & 2 \\ \hline
\hline
\end{tabular}
\label{tab:Summary}
\end{center}
\end{table*}

\subsection{Octagons}
\begin{figure}[h!bt]
  \begin{centering}
    $\begin{array}{c@{\hspace{0.5in}}c}
      \includegraphics[width=0.2\textwidth]{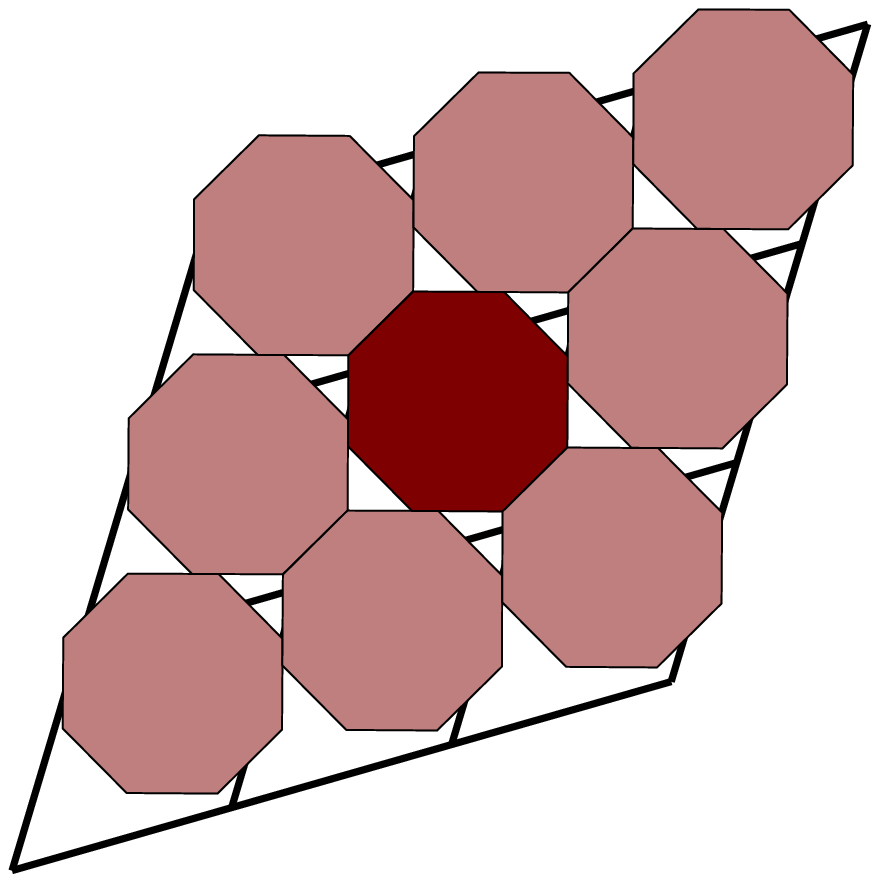}
      &
      \includegraphics[width=0.2\textwidth]{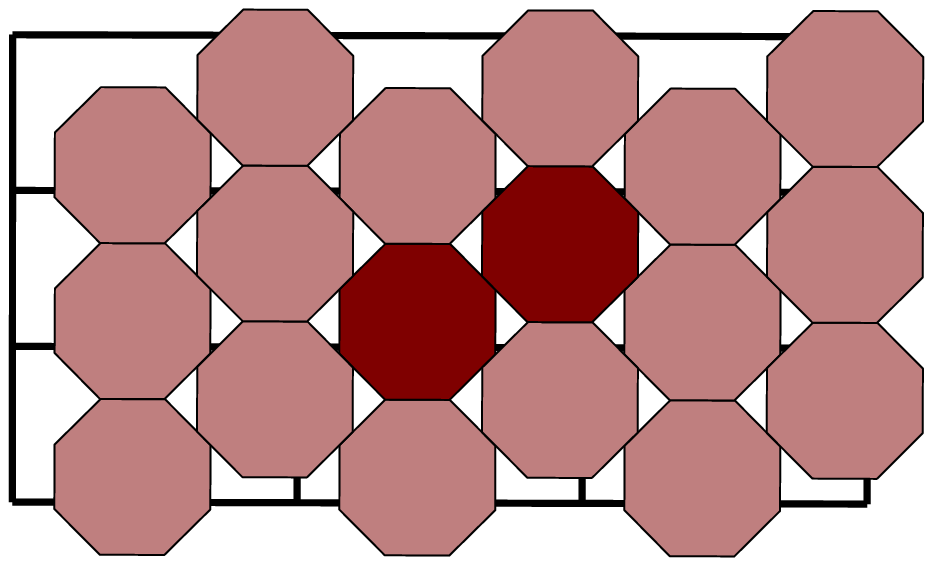}
      \\
      \mbox{(a)}
      &
      \mbox{(b)}
      \\
      \includegraphics[width=0.2\textwidth]{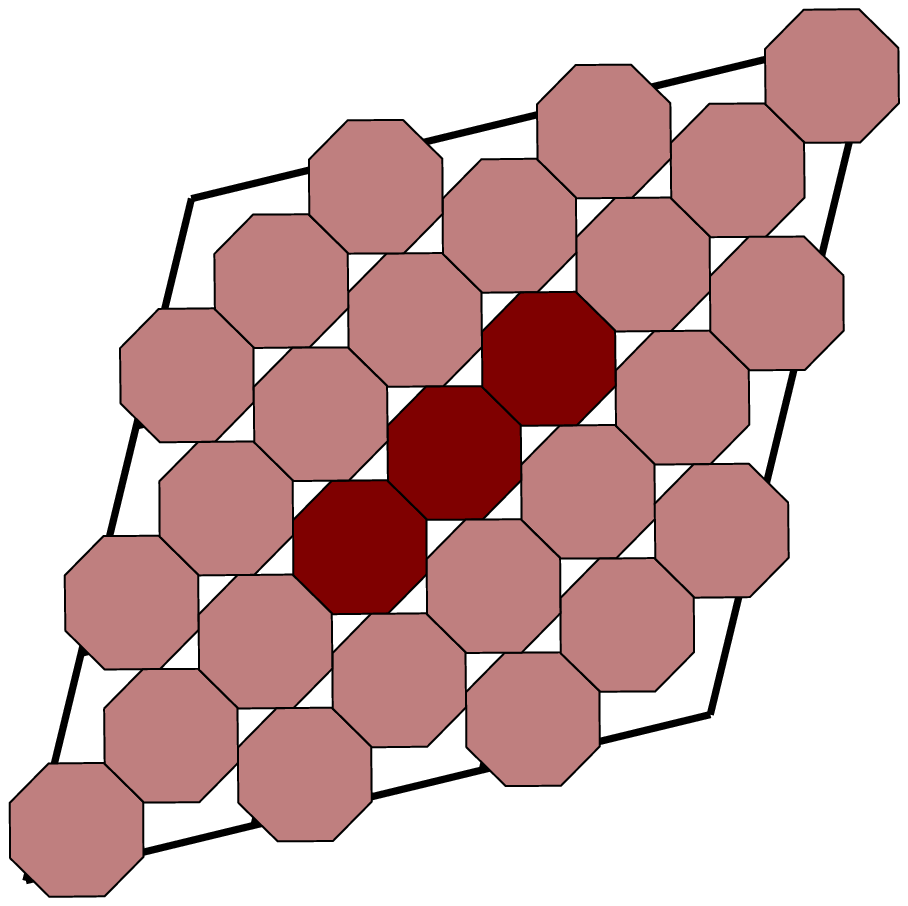}
      &
      \includegraphics[width=0.2\textwidth]{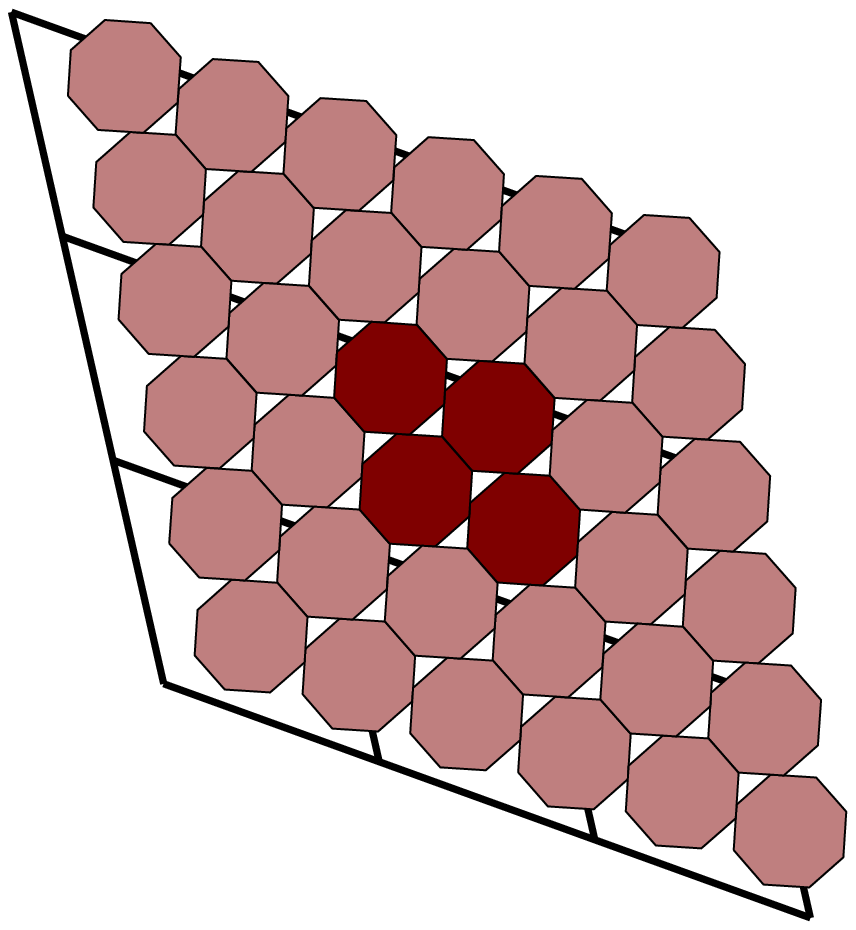}
      \\
      \mbox{(c)}
      &
      \mbox{(d)}
    \end{array}$
    \caption{(Color online) Computer-generated dense periodic packings
      of octagons using one- (a), two- (b), three- (c), and
      four-particle (d) bases. The packing densities are $\phi =
      0.906144$, $0.906029$, $0.905929$, and $0.905903$,
      respectively} 
    \label{fig:Octagons}
  \end{centering}
\end{figure}

Periodic packings of congruent copies of regular octagons were
generated using the ASC algorithm with one-, two-, three-, and
four-particle bases.  Some noteworthy computer-generated packings are
shown in Figure \ref{fig:Octagons}.  The ASC algorithm finds periodic
packings of octagons in accordance with Fejes T\'oth's well-known
theorem that the densest packing of a centrally-symmetric convex
particle shape is realized by a lattice packing \cite{FT1950}.
Moreover, the packing densities found via numerical simulation are
remarkably close to the theoretical optimum: $\delta \phi_1 = 2.0
\times 10^{-5}$, $\delta \phi_2 = 1.35 \times 10^{-4}$, $\delta \phi_3
= 2.35 \times 10^{-4}$, and $\delta \phi_4 = 2.61 \times 10^{-4}$,
where $\delta \phi_N$ denotes the difference in packing density
between the analytical optimal construction and the best numerical
result using an $N$-particle basis.

\subsection{Pentagons}
\begin{figure}[h!bt]
  \begin{centering}
    $\begin{array}{c@{\hspace{0.5in}}c}
      \includegraphics[width=0.2\textwidth]{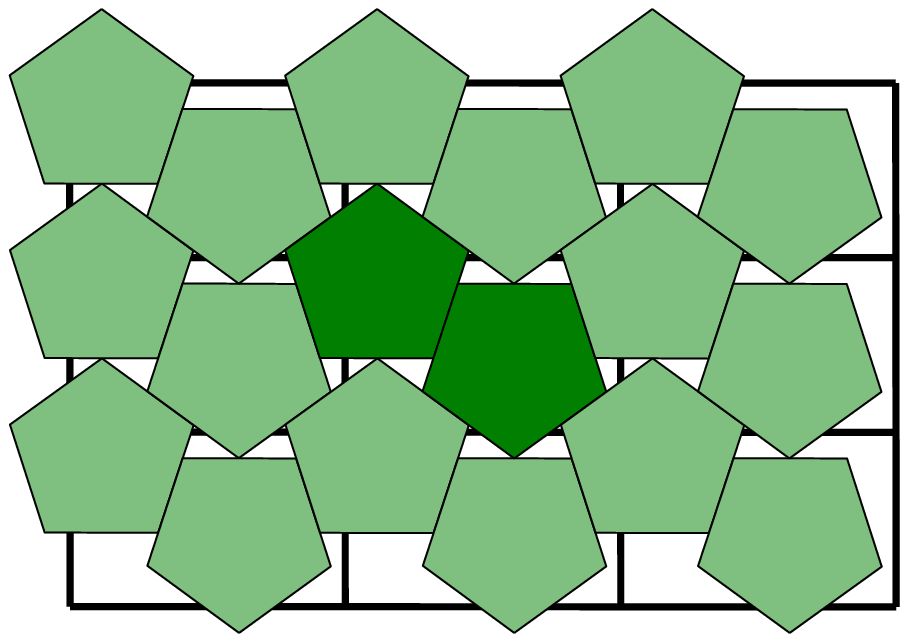}
      &
      \includegraphics[width=0.2\textwidth]{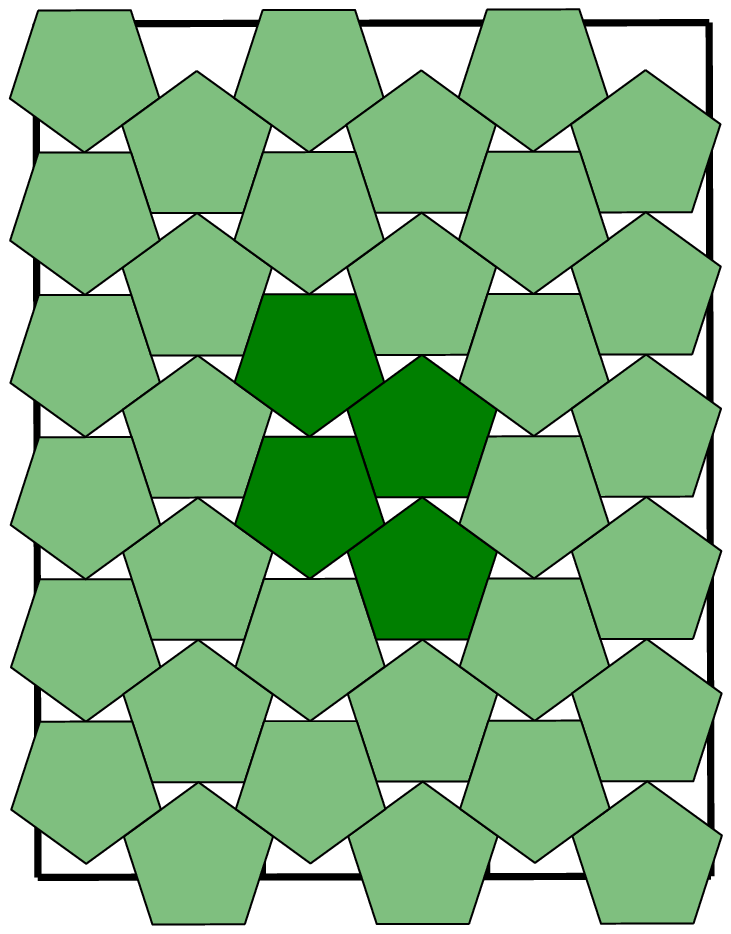}
      \\
      \mbox{(a)}
      &
      \mbox{(b)}
    \end{array}$
    \caption{(Color online) Computer-generated dense periodic packings
      of pentagons using two- (a) and four-particle (b) bases.  The
      packing densities are $\phi = 0.921102$ and $0.921301$,
      respectively.} 
    \label{fig:Pentagons}
  \end{centering}
\end{figure}

Periodic packings of congruent copies of regular pentagons were
generated using the ASC algorithm with two- and four-particle bases.
Figure \ref{fig:Pentagons} shows some noteworthy numerical findings.
These results closely resemble the best-known packing of pentagons,
whose packing density is $\phi = (5 - \sqrt{5}) / 3 = 0.921311 \dots$;
the packings shown in Figures \ref{fig:Pentagons}a and
\ref{fig:Pentagons}b differed from this by $2.09 \times 10^{-4}$ and
$1.0 \times 10^{-5}$, respectively.  These packings show the structure
conjectured to be the densest packing of pentagons
\cite{HenleyPentagons,KuperbergDoubleLattice,PentagonExperiment,FrenkelPentagons}.
Indeed, it is conjectured that, for a wide class of bodies lacking
central symmetry, the densest packing is realized in a double-lattice
configuration.  The regular pentagon is one of these shapes.

\subsection{Fat Crosses}
\begin{figure}[h!bt]
\begin{centering}
\psfrag{delta}{$w$}
\psfrag{rho}{$\phi$}
\includegraphics[width=2.8in]{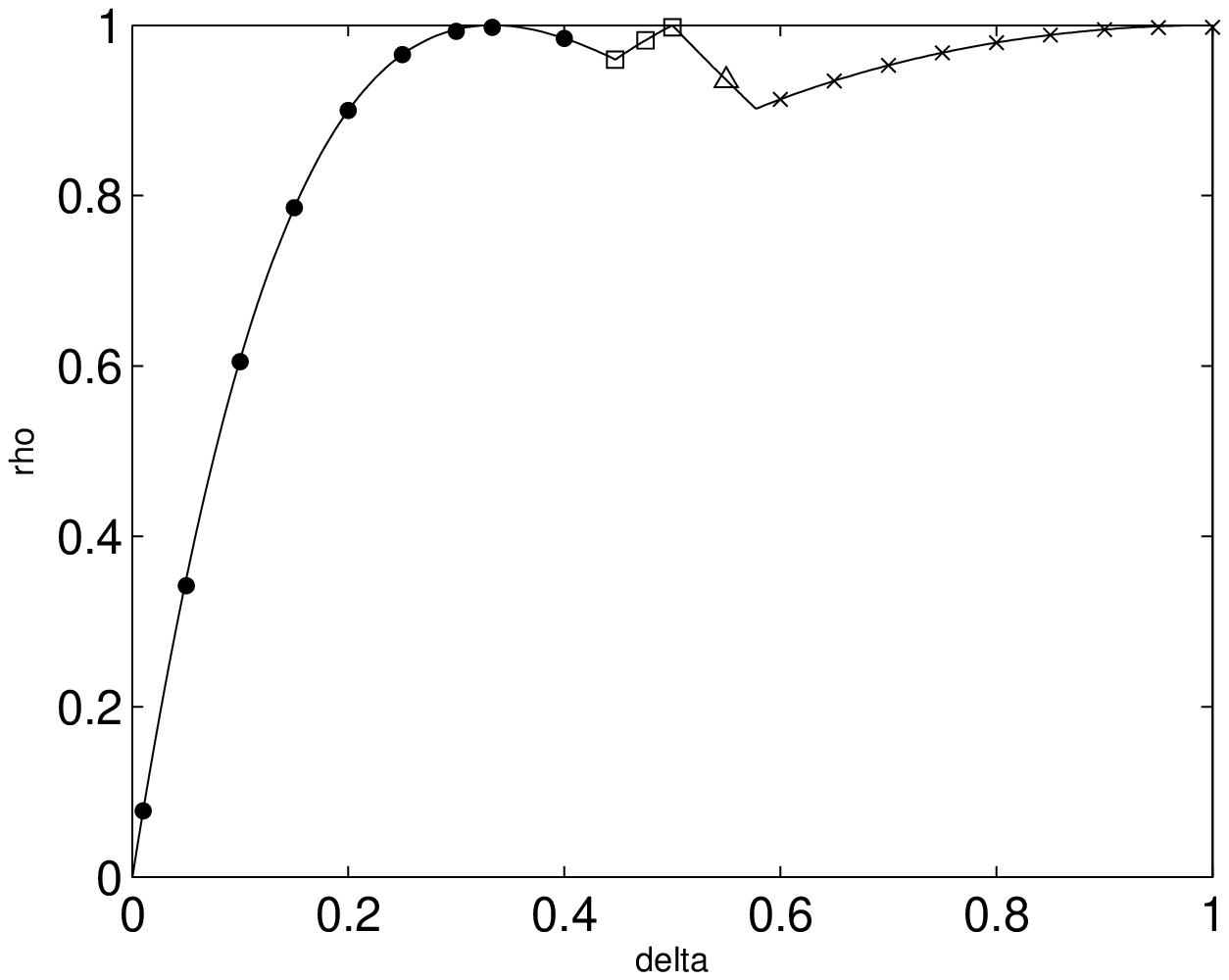}
\caption{Analytically-derived (solid curve) and computer-generated
  (data points) packing densities for packings of fat crosses as a
  function of width parameter $w$.  The structures are labeled as
  follows: filled circle = $L_1$, square = $L_2$, triangle = $L_3$,
  cross = $L_4$.}
\label{fig:CrossPlot}
\end{centering}
\end{figure}

\begin{figure}[h!b!t!]
  \begin{centering}
    $\begin{array}{c@{\hspace{0.3in}}c@{\hspace{0.3in}}c}
      \includegraphics[width=0.23\textwidth]{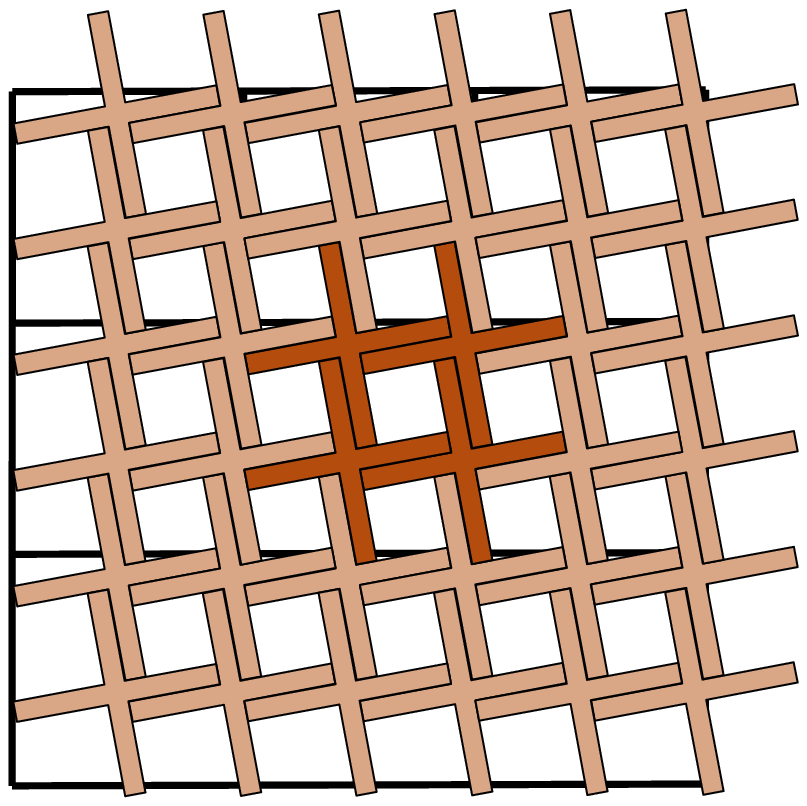}
      &
      \includegraphics[width=0.23\textwidth]{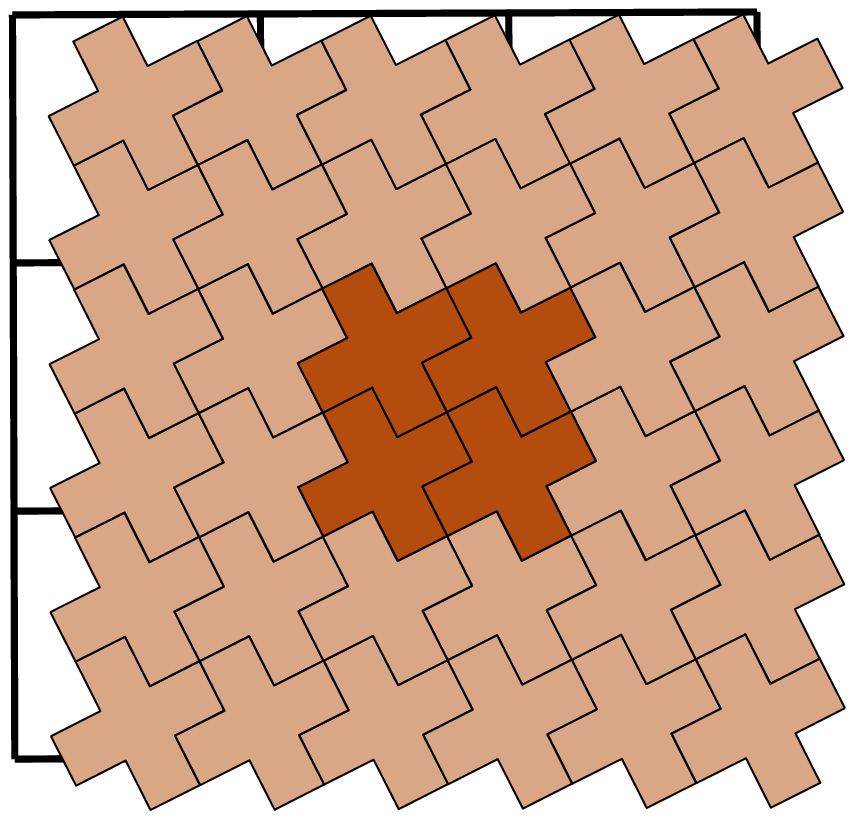}
      &
      \includegraphics[width=0.23\textwidth]{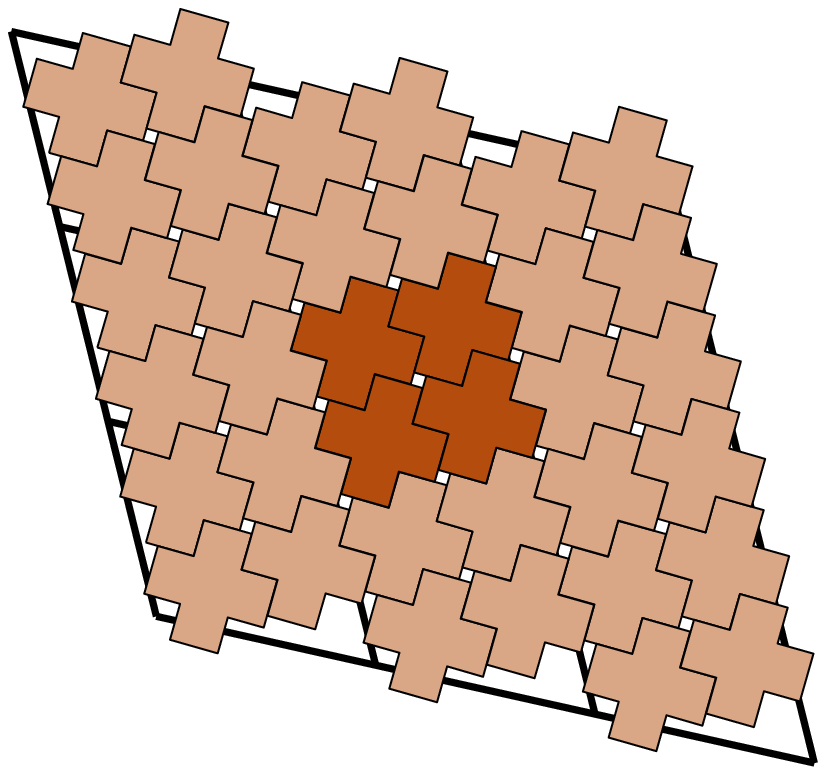}
      \\
      \mbox{(a)}
      &
      \mbox{(b)}
      &
      \mbox{(c)}
      \\
      \includegraphics[width=0.23\textwidth]{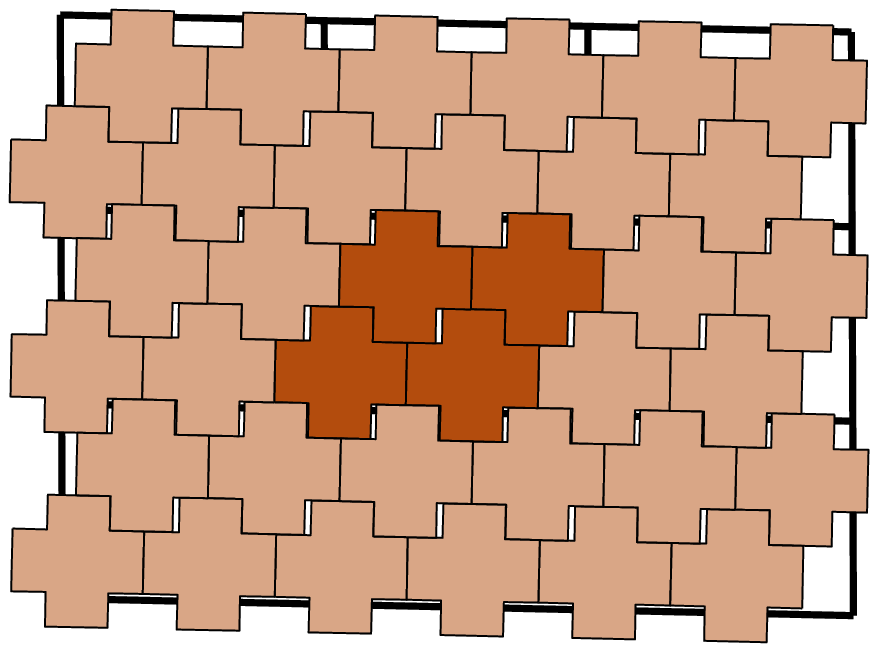}
      &
      \includegraphics[width=0.23\textwidth]{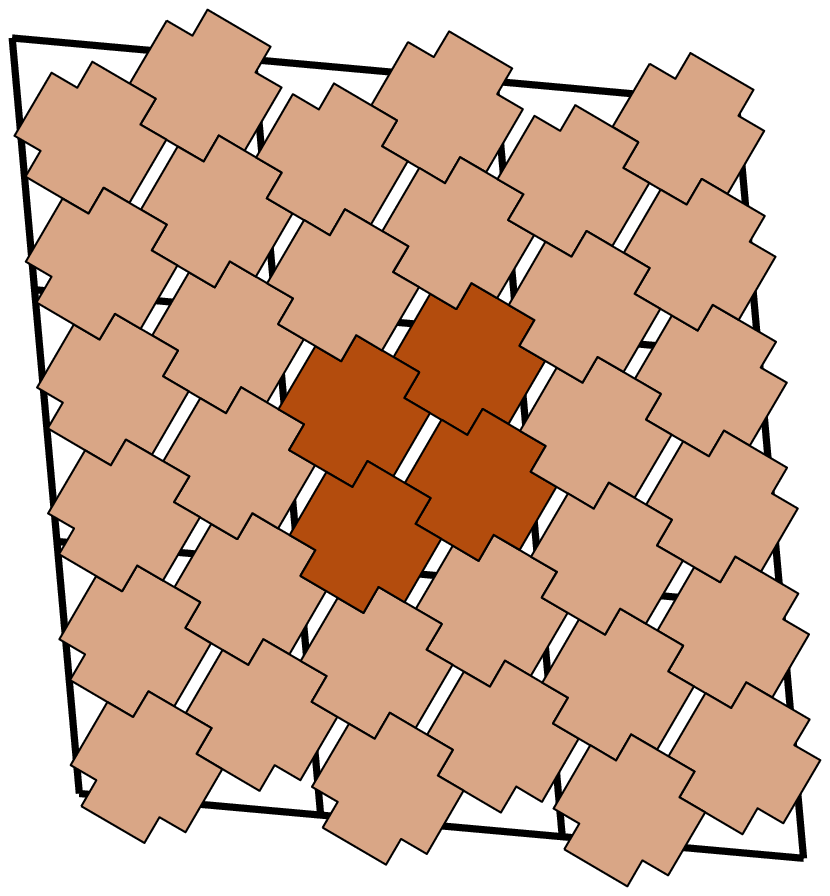}
      &
      \includegraphics[width=0.23\textwidth]{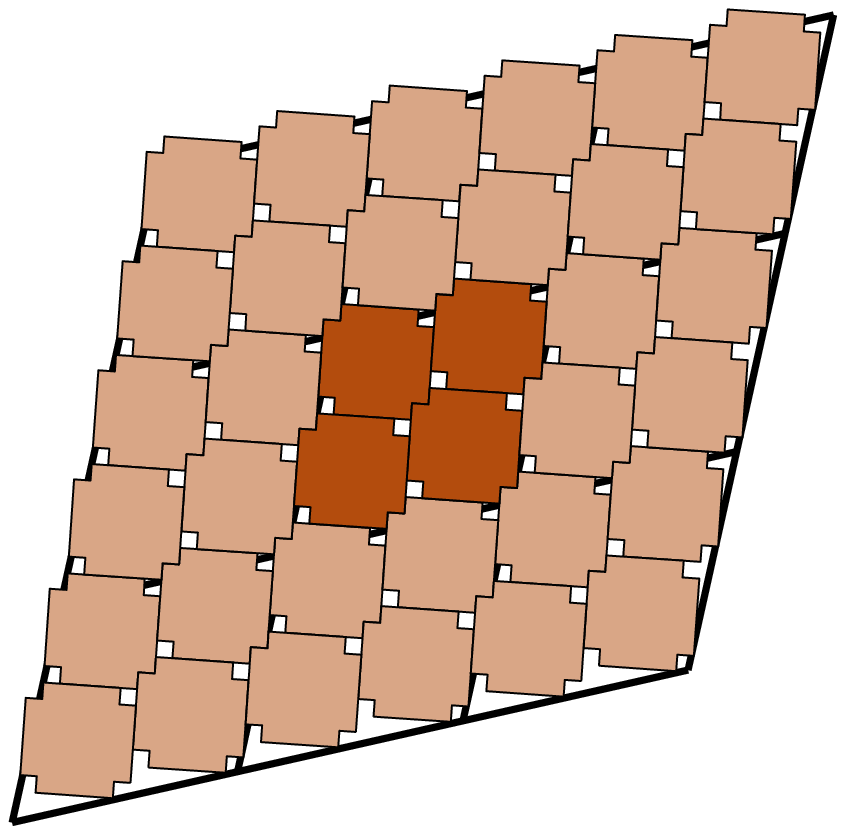}
      \\
      \mbox{(d)}
      &
      \mbox{(e)}
      &
      \mbox{(f)}
    \end{array}$
    \caption{(Color online) Computer-generated dense periodic packings
      of fat crosses for various values of $w$ showing the four
      structures observed.  The $L_1$ structure is shown for $w=1/10$ (a),
      $1/3$ (b), and $2/5$ (c); the $L_2$ structure is shown for
      $w=19/40$ (d); the $L_3$ structure is shown for $w=11/20$ (e);
      and the $L_4$ structure is shown for $w=7/10$ (f).}
    \label{fig:Crosses}
  \end{centering}
\end{figure}

Periodic packings of congruent copies of fat crosses were generated
using the ASC algorithm with four-particle bases for various values of
$w$.  From these simulations, four different packing behaviors were
observed.  Figure \ref{fig:CrossPlot} compares the packing density of
a number of computer-generated cases against the analytical putative
maximum packing density.  The curve contains four piecewise-smooth
regimes, corresponding to four different structures, identified as
$L_1$, $L_2$, $L_3$, and $L_4$, in order of increasing $w$.  Figure
\ref{fig:Crosses} shows a few selected packings to show these
different configurations.  Notice that at $w = 1 / 3, 1 / 2,~{\rm
  and}~1$, the packing becomes a tiling.  It is interesting to note
that, though this shape is concave, all of the densest packings found
are lattice packings.  Another interesting property of these packings
is that, in the limit case $w=0$, the fat cross is equivalent (up to a
scaling constant) to the limit case of a superdisk, $\lim \limits_{p
  \rightarrow 0} \{(x,y) \in \reals^2:~|x|^{2p} + |y|^{2p} \le 1\}$,
and the analytical packing structure matches this superdisk's
already-known densest packing \cite{YangSuperdisks}.  The fat crosses'
putative maximum packing density is tabulated for several
representative values in Table \ref{tab:Cross}; see the {\it
Supplemental Material} for mathematical details about the various
packing structures.

\begin{table}[h!t]
\begin{center}
\caption{Putative maximum packing densities of congruent copies of fat
  crosses for selected values of $w$.}
\begin{tabular}{lcrl}
\hline \hline
$w$   & \hspace{42 mm} & & $\phi(w)$ \\ \hline
0.1   & \hspace{20 mm} & $76 / 125 $ & $ = 0.608$ \\
0.2   & \hspace{20 mm} & $9  / 10  $ & \\ 
0.3   & \hspace{20 mm} & $204 / 205$ & $ = 0.995121 \dots$ \\
0.4   & \hspace{20 mm} & $64  / 65 $ & $ = 0.984615 \dots$ \\
0.48  & \hspace{20 mm} & $912 / 925$ & $ = 0.985945 \dots$ \\
0.55  & \hspace{20 mm} & $29  / 31 $ & $ = 0.935483 \dots$ \\
0.6   & \hspace{20 mm} & $21  / 23 $ & $ = 0.913043 \dots$ \\
0.7   & \hspace{20 mm} & $182 / 191$ & $ = 0.952879 \dots$ \\
0.8   & \hspace{20 mm} & $48  / 49 $ & $ = 0.979591 \dots$ \\
0.9   & \hspace{20 mm} & $198 / 199$ & $ = 0.994974 \dots$ \\ \hline \hline
\end{tabular}
\label{tab:Cross}
\end{center}
\end{table}

\subsection{Curved Triangles}
\begin{figure}[hbt]
\begin{centering}
\psfrag{k}{$k$}
\psfrag{rho}{$\phi$}
\includegraphics[width=2.8in]{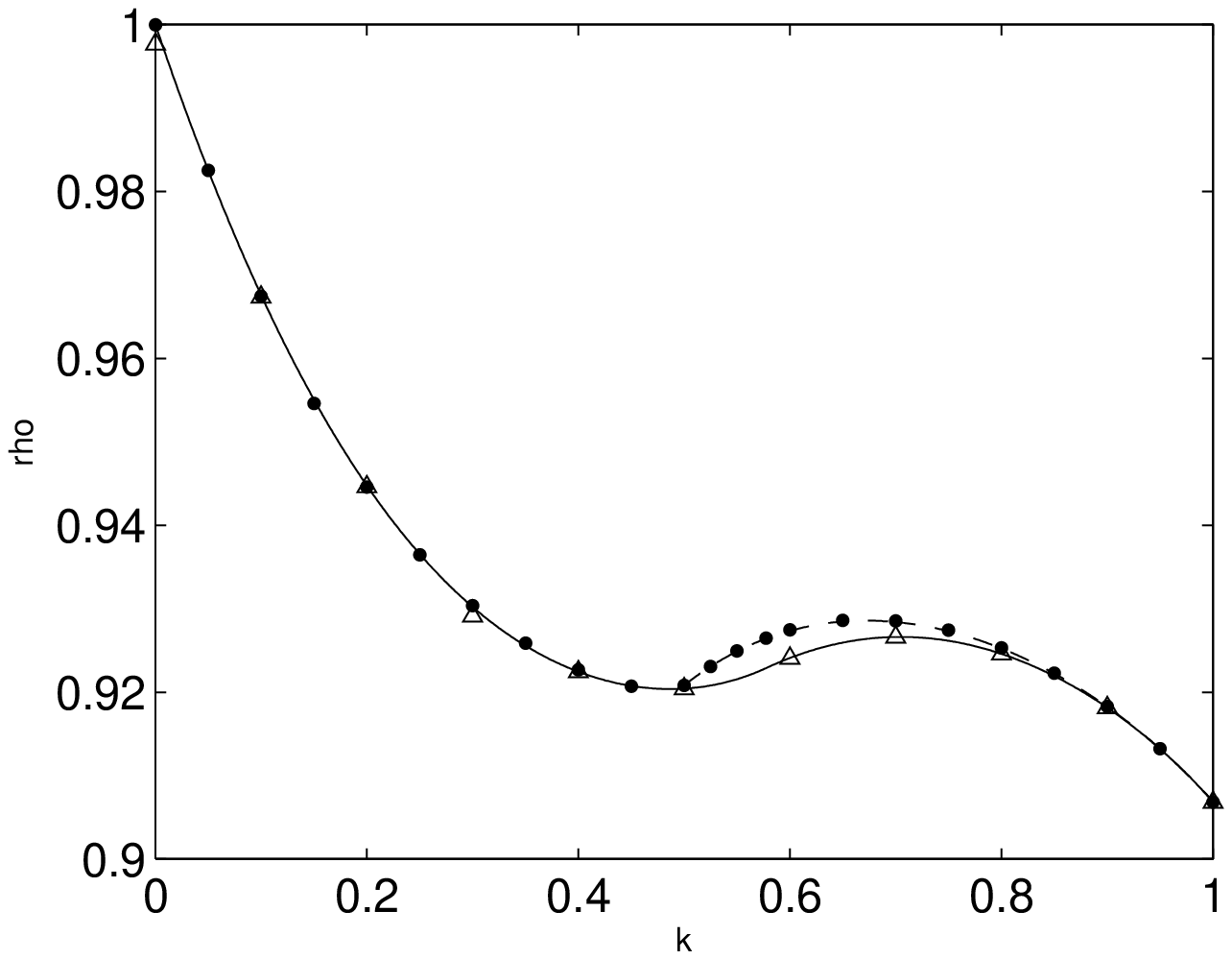}
\caption{Analytically-derived and computer-generated packing densities
  of curved triangles for various values of $k$.  The solid curve
  denotes the density of the optimal double lattice packing, and the
  dashed curve shows the density of the optimal periodic packing with
  a four-particle basis; triangle and filled circle data points denote
  densities obtained by computer-generated packings with two-
  and four-particle bases, respectively.}
\label{fig:TrianglePlotConvex}
\end{centering}
\end{figure}

Periodic packings of congruent copies of convex curved triangles were
generated using the ASC algorithm with two- and four-particle bases
for various values of $k$.  Using a method of Kuperberg and Kuperberg
\cite{KuperbergDoubleLattice}, the densest double-lattice packing of
convex curved triangles was derived as a function of the curvature
parameter $k$.  This function is plotted as a solid curve in Figure
\ref{fig:TrianglePlotConvex}, and computer-generated packings with a
two-particle basis are plotted as triangular marks; it was observed
that the results predicted by the Monte Carlo method using a
two-particle basis closely matched the optimal double-lattice packing
for all of the chosen values of $k$.

\begin{figure}[h!bt]
  \begin{centering}
    $\begin{array}{c@{\hspace{0.5in}}c}
      \includegraphics[width=0.2\textwidth]{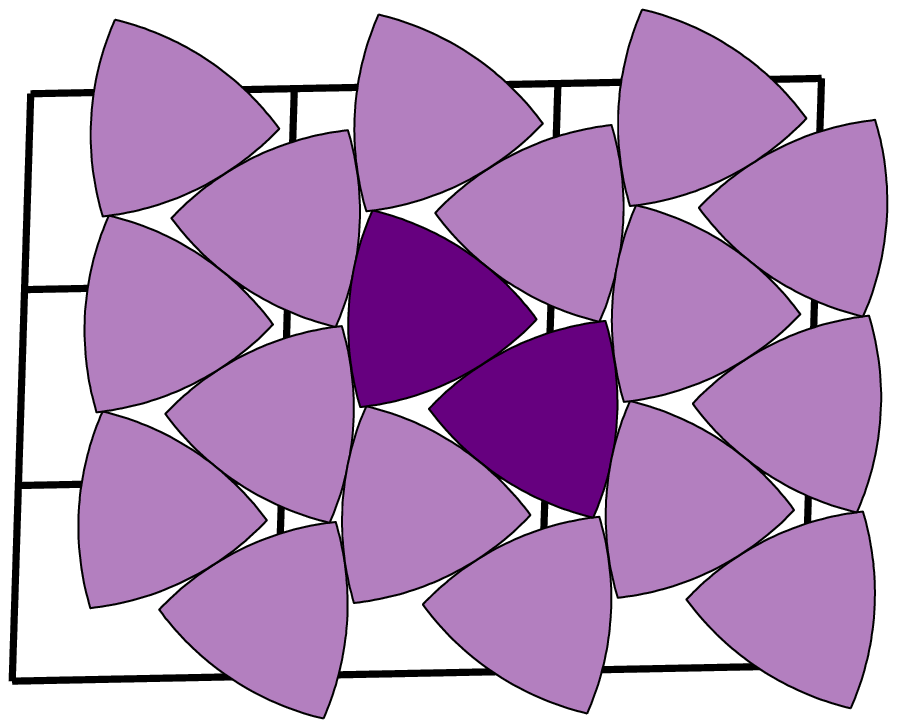}
      &
      \includegraphics[width=0.2\textwidth]{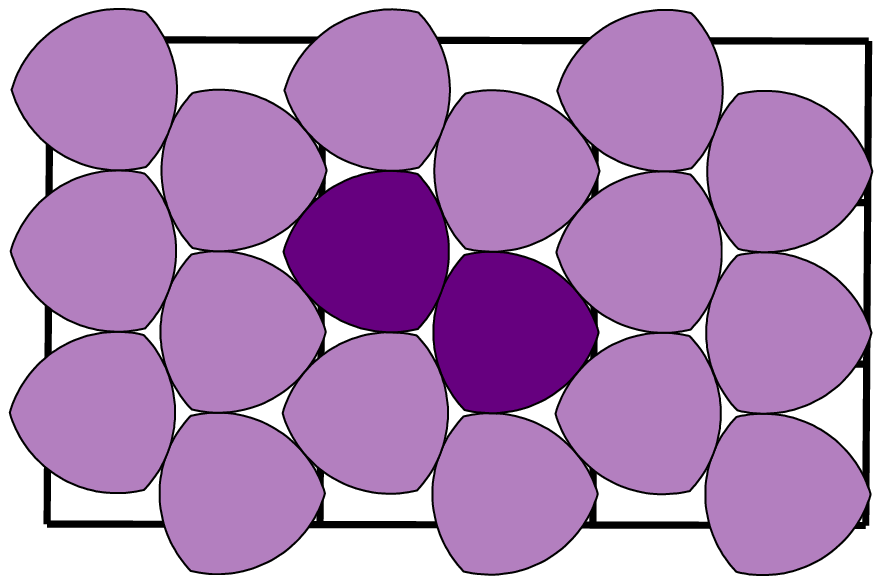}
      \\
      \mbox{(a)}
      &
      \mbox{(b)}
    \end{array}$
    \caption{(Color online) Computer-generated dense periodic packings
      of curved triangles with a two-particle basis for $k = 2 / 5$
      (a) and $4/5$ (b); $\phi = 0.922437$ and $0.924589$, respectively.}
    \label{fig:CurvedTriangle2Basis}
  \end{centering}
\end{figure}

\begin{figure}[h!bt]
  \begin{centering}
    $\begin{array}{c@{\hspace{0.5in}}c}
      \includegraphics[width=0.2\textwidth]{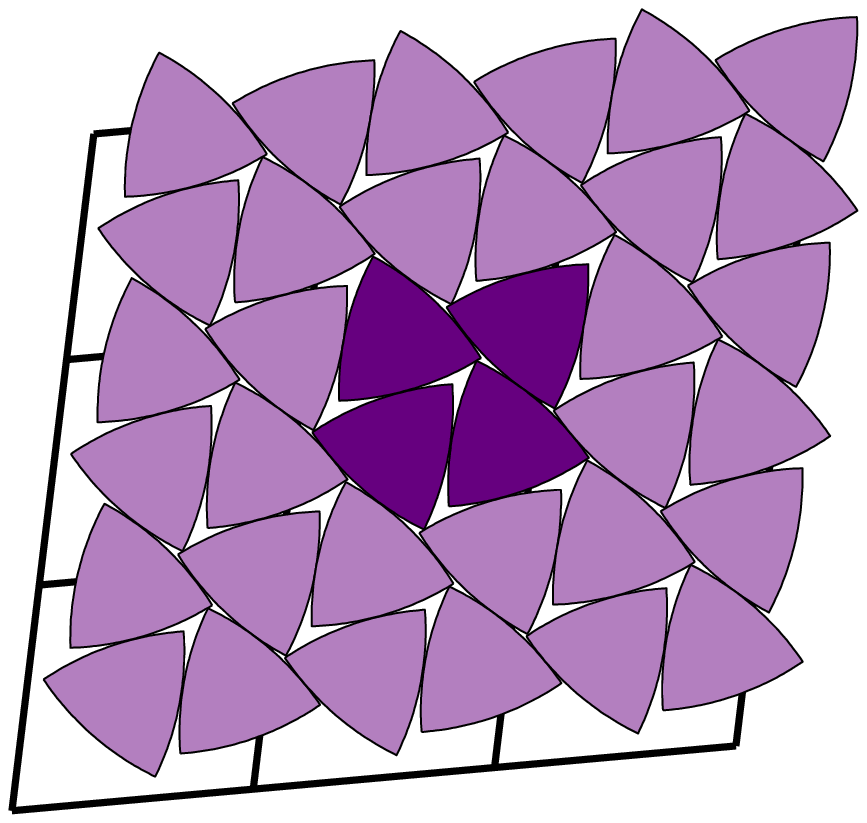}
      &
      \includegraphics[width=0.2\textwidth]{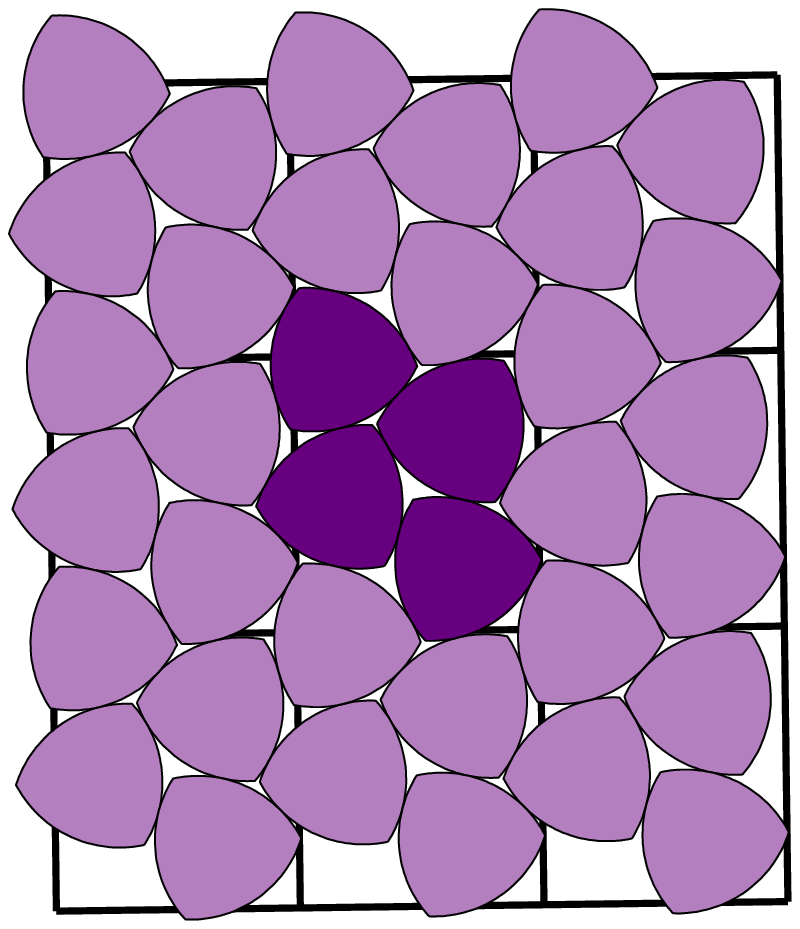}
      \\
      \mbox{(a)}
      &
      \mbox{(b)}
    \end{array}$
    \caption{(Color online) Computer-generated dense periodic packings
      of curved triangles with a four-particle basis for $k = 3 / 10$
      (a) and $7/10$ (b); $\phi = 0.921528$ and $0.921460$,
      respectively.}
    \label{fig:CurvedTriangle4Basis}
  \end{centering}
\end{figure}

Figures \ref{fig:CurvedTriangle2Basis} and
\ref{fig:CurvedTriangle4Basis} show a few of the packings that were
generated by the ASC algorithm with two- and four-particle bases,
respectively.  Notice that the densest packings with a two-particle
basis are double-lattice packings, implying a centrally-symmetric
basis.  However, in all of the cases where $0 < k < 1 / \sqrt{3}$, it
was observed that the triangles do {\it not} line up such that their
vertices touch.  Instead, the two adjacent triangles are rotated
slightly so that the vertex of one contacts the arc of the other.
Indeed, when determining the optimal double-lattice structure by a
method of Kuperberg and Kuperberg \cite{KuperbergDoubleLattice}, it
was found that this small perturbation actually increases the packing
density when $0 < k < 1 / \sqrt{3}$, meaning that this characteristic
of the numerical results is not due to numerical inaccuracies, but, on
the contrary, showcases the sensitivity of the stochastic search in
finding dense packings.

\begin{figure}[hbt]
\begin{centering}
\includegraphics[width=3in]{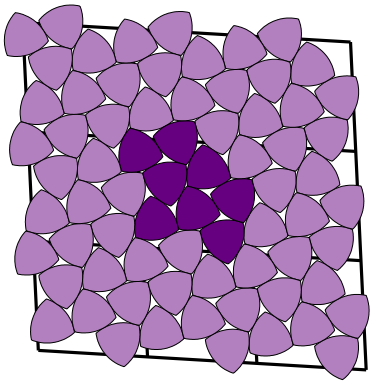}
\caption{(Color online) A computer-generated densest packing of curved
  triangles with an eight-particle basis ($k = 0.62$), showing a
  structure that is a degenerate case of the best-known four-particle
  basis.}
\label{fig:Triangle8}
\end{centering}
\end{figure}

Numerical results using a four-particle basis showed that, for
sufficiently large $k$, a packing structure exists whose density is
significantly higher than the best double-lattice packing; below this
point, the four-particle basis' optimal configuration is a degenerate
case of the two-particle basis, as shown in Figure
\ref{fig:CurvedTriangle4Basis}a; an example of the non-degenerate
four-particle basis structure is given in Figure
\ref{fig:CurvedTriangle4Basis}b.  Numerical results using an
eight-particle basis, an example of which is shown in Figure
\ref{fig:Triangle8}, provide additional evidence that the
four-particle basis is in fact the densest configuration.  These
results were verified by determining the analytical structure of the
four-particle basis that is currently the best-known packing structure
for this shape.  The densities of computer-generated packings of
curved triangles using a four-particle basis are plotted as filled
circles, and the corresponding analytical packing density is displayed
as a dashed line in Figure \ref{fig:TrianglePlotConvex}. Though
periodic tilings of convex particles have been shown that exhibit more
than a two-particle basis
\cite{ReinhardtThesis,PentagonTilings2,Heesch4Basis}, we do not know
of any other convex, non-tiling shapes that have been observed to
exhibit this behavior \footnote{We do not consider particle shapes
  derived from applying trivial perturbations (that is, those that do
  not alter the densest packing structure) to shapes that form
  tilings}.

\begin{figure}[h!bt]
\begin{centering}
\psfrag{A}{$A_0$}
\psfrag{B}{$B_0$}
\psfrag{C}{$C_0$}
\psfrag{D}{$D_0$}
\psfrag{A1}{$A_1$}
\psfrag{B1}{$B_1$}
\psfrag{C1}{$C_1$}
\psfrag{D1}{$D_1$}
\psfrag{A2}{$A_2$}
\psfrag{B2}{$B_2$}
\psfrag{C2}{$C_2$}
\psfrag{D2}{$D_2$}
\includegraphics[width=0.4\textwidth]{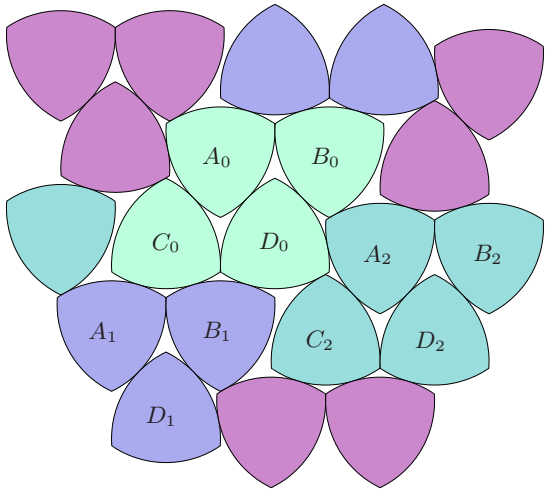}
\caption{(Color online) A portion of the analytical packing without
  any applied rotation ($k = 0.65$).  Notice that a gap exists between
  $B_1$ and $C_2$ (and their periodic images).  Applying the rotation
  closes this gap and maximizes the packing density.}
\label{fig:4BodyNoRotate}
\end{centering}
\end{figure}

\begin{figure*}[h!bt]
  \begin{centering}
    \psfrag{A0}[c][c]{$A_0$}
    \psfrag{B0}[c][c]{$B_0$}
    \psfrag{C0}[c][c]{$C_0$}
    \psfrag{D0}[c][c]{$D_0$}
    \psfrag{A1}[c][c]{$A_1$}
    \psfrag{B1}[c][c]{$B_1$}
    \psfrag{C1}[c][c]{$C_1$}
    \psfrag{D1}[c][c]{$D_1$}
    \psfrag{A2}[c][c]{$A_2$}
    \psfrag{B2}[c][c]{$B_2$}
    \psfrag{C2}[c][c]{$C_2$}
    \psfrag{D2}[c][c]{$D_2$}
    $\begin{array}{c@{\hspace{0.3in}}c}
      \includegraphics[width=0.46\textwidth]{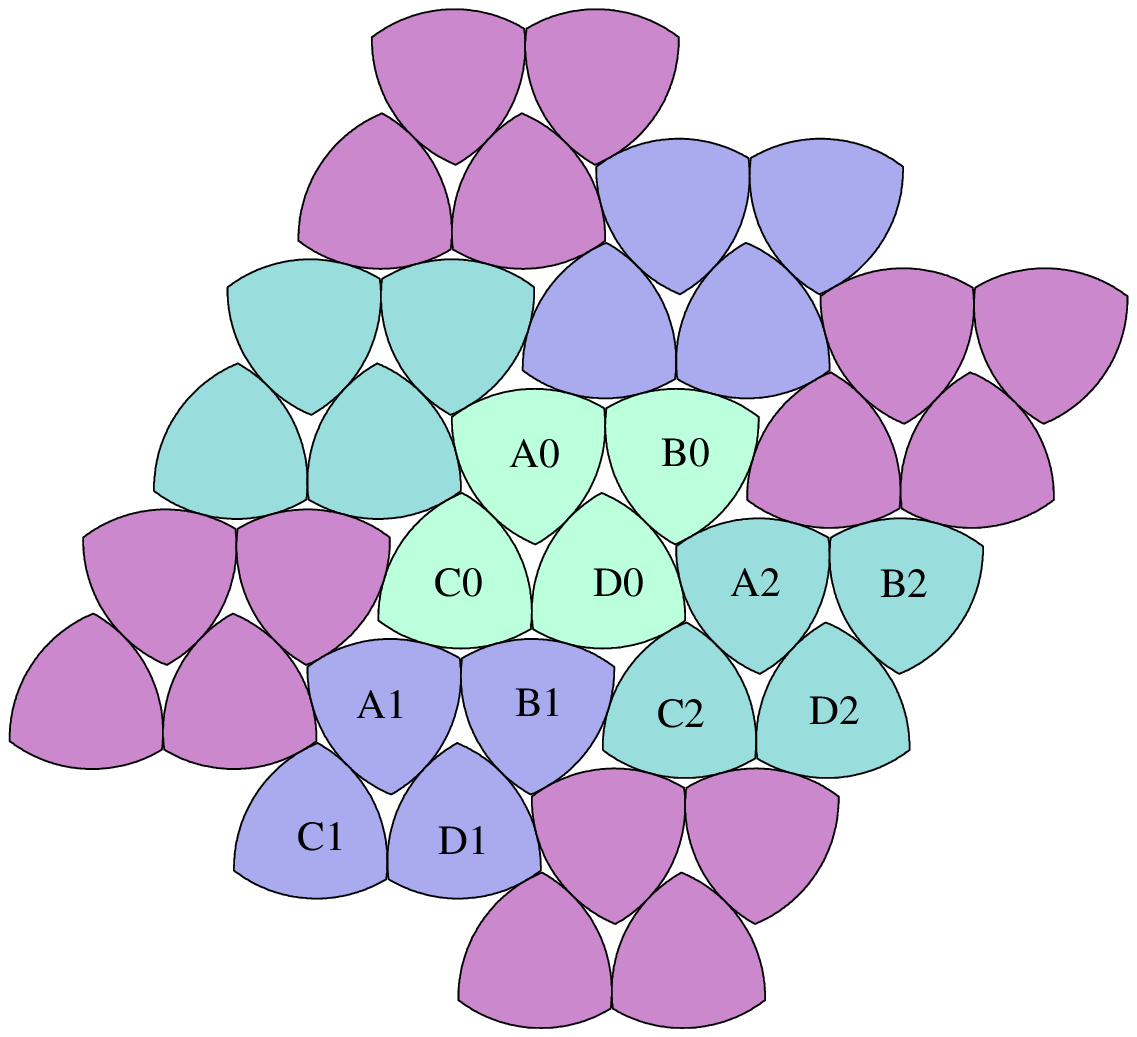}
      &
      \psfrag{A}{$A_0$}
      \psfrag{B}{$B_0$}
      \psfrag{C}{$C_0$}
      \psfrag{D}{$D_0$}
      \includegraphics[width=0.43\textwidth]{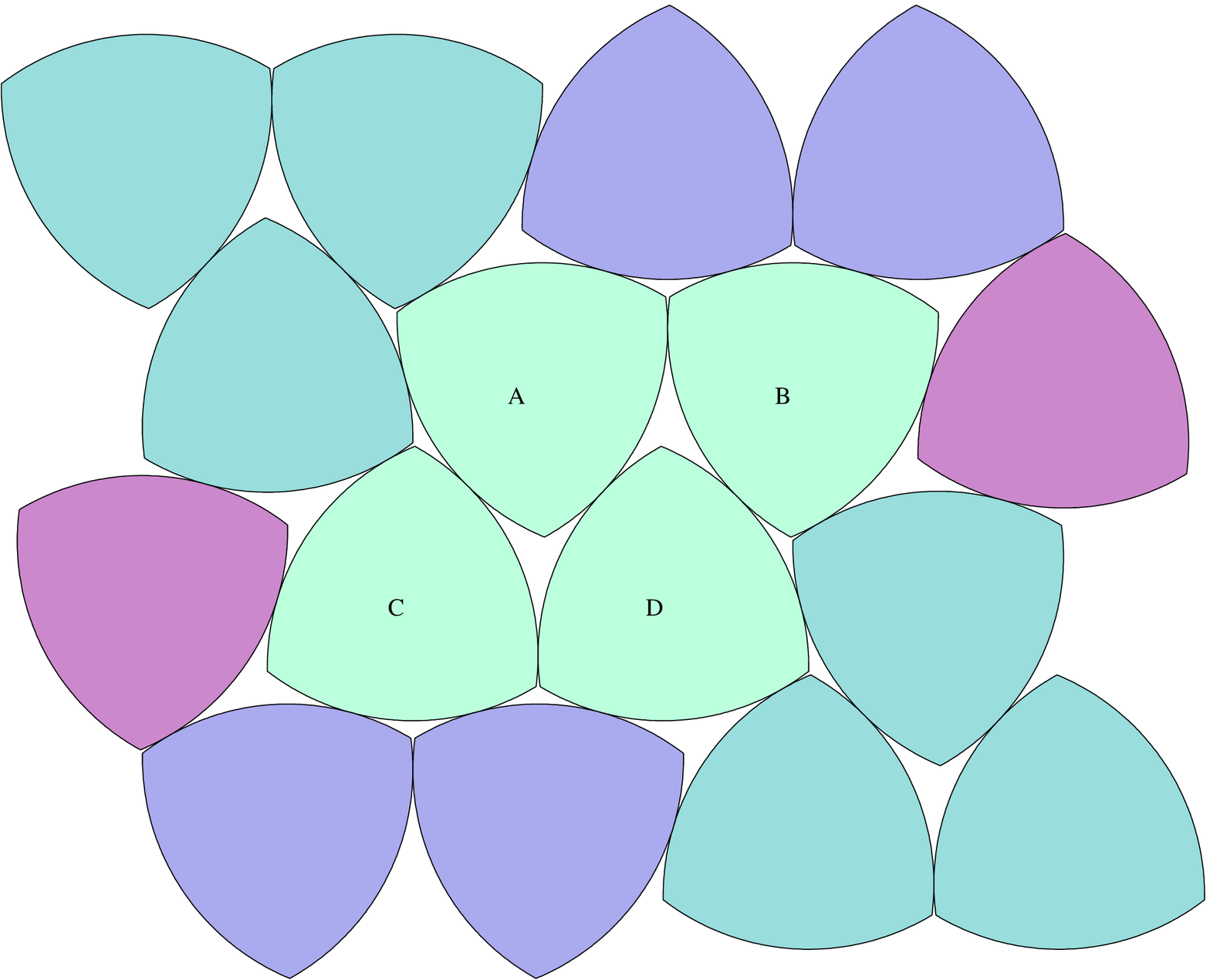}
      \\
      \mbox{(a)}
      &
      \mbox{(b)}
    \end{array}$
    \caption{(Color online) The analytically-derived packing of curved
      triangles ($k = 0.65$) with the applied rotation (a); a
      zoomed-in view is provided in (b).  Triangles
      $A_0$ and $D_0$ are rotated anti-clockwise by $\theta$, and
      triangles $B$ and $C$ are rotated clockwise by $\theta$, where
      $\theta = 0.057~{\rm rad}$.  $\phi = 0.928473\dots$.}
    \label{fig:CS4BasisAnalytical}
  \end{centering}
\end{figure*}

In order to create this four-particle packing structure, the particles
in the FC, denoted $A_0$, $B_0$, $C_0$, and $D_0$, are placed facing
straight-up and straight-down, where particles $A_0$ and $D_0$ are
both in contact with all of the other particles in the basis, as shown
in Figure \ref{fig:4BodyNoRotate}.  The lattice vectors
$\vc{\lambda}_1$ and $\vc{\lambda}_2$ are found by finding the
positions of shapes $A_1$ and $A_2$ (both of which are periodic images
of $A_0$) relative to $A_0$.  All of the bodies in the basis are
rotated by the same angle $\theta$: $A$ and $D$ are rotated
anti-clockwise, and $B$ and $C$ are rotated clockwise.  Doing this
causes $B_1$ and $C_2$ (and their periodic images) to come into
contact.  Once this contact is established, the packing achieves its
maximum density; this result is shown in Figure
\ref{fig:CS4BasisAnalytical}.  The curved triangles' putative maximum
packing density is tabulated for some representative values of $k$ in
Table \ref{tab:Triangle}; see the {\it Supplemental Material} for
details of the analytical construction of this packing.  In addition,
computer-generated packings of {\it concave} curved triangles are
provided in the Appendix without analytical constructions; all of
these packings appear to be double-lattice packings.

\begin{table}[h!t]
\begin{center}
\caption{Putative maximum packing densities of congruent copies of
  convex curved triangles for selected values of $k$.}
\begin{tabular}{lcc}
\hline \hline
$k$ & \hspace{60 mm} & $\phi(k)$ \\ \hline
0.1 && $0.967582 \dots$ \\ 
0.2 && $0.944761 \dots$ \\ 
0.3 && $0.930217 \dots$ \\ 
0.4 && $0.922458 \dots$ \\ 
0.5 && $0.922458 \dots$ \\ 
0.6 && $0.927362 \dots$ \\ 
0.7 && $0.928415 \dots$ \\ 
0.8 && $0.925249 \dots$ \\ 
0.9 && $0.918262 \dots$ \\ \hline \hline
\end{tabular}
\label{tab:Triangle}
\end{center}
\end{table}

\subsection{Moon-like Shapes}
\begin{figure}[hbt]
\begin{centering}
\psfrag{rho}{$\phi$}
\psfrag{k}{$k$}
\includegraphics[width=3in]{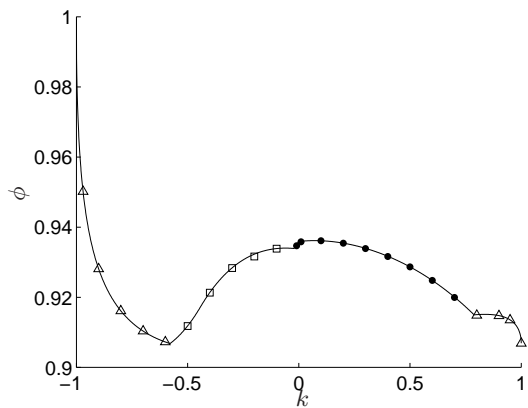}
\caption{Putative maximum packing densities for congruent copies of
  moon-like shapes as a function of $k$.  The three different
  structures are denoted as follows: triangle = $D_1$, square = $B$,
  dot = $D_2$); the solid curve shows the analytical densities.}
\label{fig:MoonDensities}
\end{centering}
\end{figure}

\begin{figure}[hbt]
  \begin{centering}
    $\begin{array}{c@{\hspace{0.5in}}c}
      \includegraphics[width=0.2\textwidth]{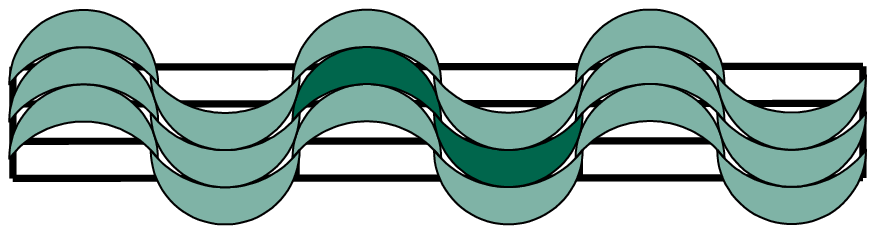}
      &
      \includegraphics[width=0.2\textwidth]{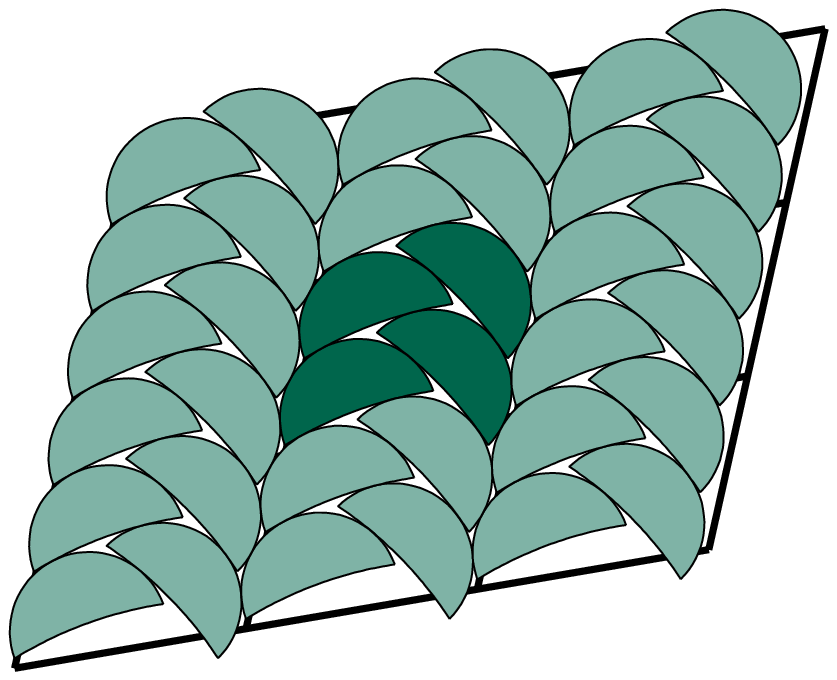}
      \\
      \mbox{(a)}
      &
      \mbox{(b)}
      \\
      \includegraphics[width=0.2\textwidth]{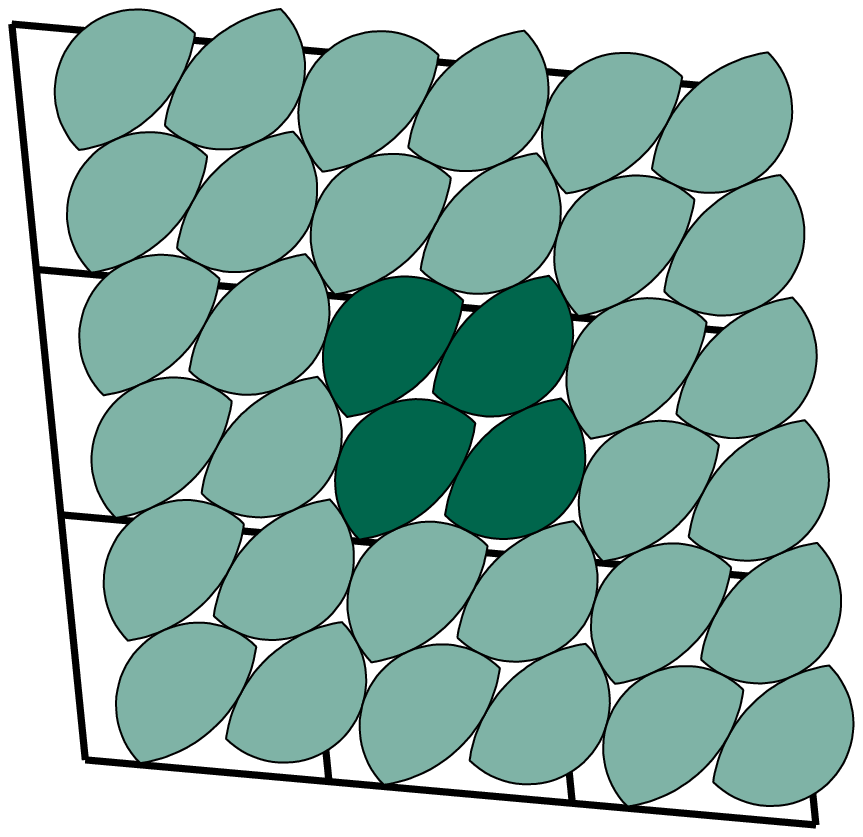}
      &
      \includegraphics[width=0.2\textwidth]{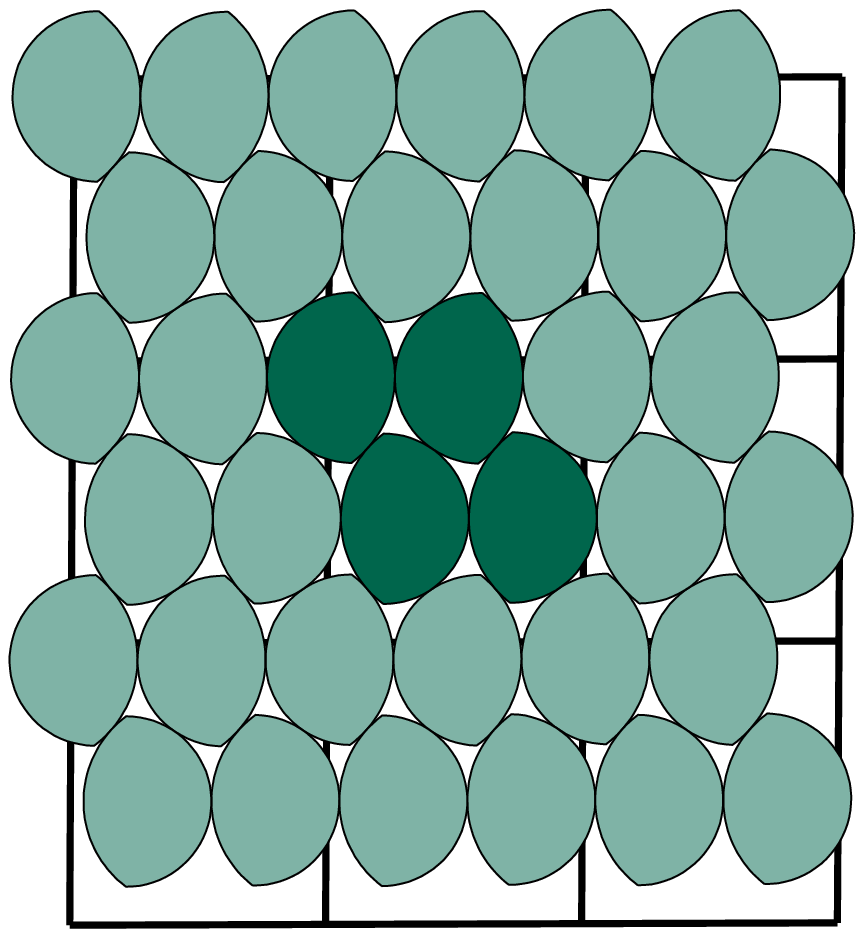}
      \\
      \mbox{(c)}
      &
      \mbox{(d)}
    \end{array}$
    \caption{(Color online) Computer-generated dense periodic packings
      of moon-like shapes for various values of $k$, displaying the
      three different packing structures observed.  (a): $k = -0.8$
      ($D_1$-structure, crescent instance).  (b): $k = -0.2$
      ($B$-structure).  (c): $k = 0.6$ ($D_2$-structure).  (d): $k =
      0.9$ ($D_1$-structure, gibbous instance).}
    \label{fig:MoonPackings}
  \end{centering}
\end{figure}

Periodic packings of congruent copies of moon-like shapes were
generated using the ASC algorithm with two- and four-particle bases
for various values of $k$.  The densest computer generated packings
are plotted as data points along solid curves denoting three different
analytical constructions in Figure \ref{fig:MoonDensities}.  The first
structure observed was a double-lattice packing shown in Figures
\ref{fig:MoonPackings}a and \ref{fig:MoonPackings}d.  This structure is
indicated in Figure \ref{fig:MoonDensities} by triangular data points
and will be referred to as the $D_1$ structure.  The second structure
observed was a non-double-lattice periodic packing with a two-particle
basis, shown in Figure \ref{fig:MoonPackings}b; it is indicated in
Figure \ref{fig:MoonDensities} by square data points and will be
referred to as the $B$-structure.  The third structure observed was a
second double-lattice packing, shown in Figure
\ref{fig:MoonPackings}c.  This structure is indicated in Figure
\ref{fig:MoonDensities} by filled circle data points and is denoted
as the $D_2$-structure.  The putative densest packings of moon-like
shapes are tabulated as a function of $k$ in Table \ref{tab:Moon}. 

The $B$-structure is a unique structure in that its fundamental basis
possesses no inherent symmetries.  Furthermore, its discovery
underscores the utility of the ASC algorithm in determining the
densest packings that are not obvious by inspection.

\begin{table}[h!t]
\begin{center}
\caption{Putative maximum packing densities of congruent copies of
  moon-like shapes for selected values of $k$.}
\begin{tabular}{lcccccc}
\hline \hline
$k$     & \hspace{10 mm} & $\phi(k)$        & \hspace{10 mm} & $k$  & \hspace{10 mm} & $\phi(k)$ \\ \hline
$-0.9$  &                & $0.928125 \dots$ &                & 0.1  &                & $0.936131 \dots$ \\
$-0.8$  &                & $0.916116 \dots$ &                & 0.2  &                & $0.935451 \dots$ \\
$-0.7$  &                & $0.910371 \dots$ &                & 0.3  &                & $0.933966 \dots$ \\
$-0.6$  &                & $0.907259 \dots$ &                & 0.4  &                & $0.931709 \dots$ \\
$-0.5$  &                & $0.911837 \dots$ &                & 0.5  &                & $0.928675 \dots$ \\
$-0.4$  &                & $0.921385 \dots$ &                & 0.6  &                & $0.924819 \dots$ \\
$-0.3$  &                & $0.928392 \dots$ &                & 0.7  &                & $0.920053 \dots$ \\
$-0.2$  &                & $0.932303 \dots$ &                & 0.8  &                & $0.915055 \dots$ \\
$-0.1$  &                & $0.933906 \dots$ &                & 0.9  &                & $0.914783 \dots$ \\
$ 0$    &                & $0.935931 \dots$ &                & 0.95 &                & $0.913563 \dots$ \\ \hline \hline
\end{tabular}
\label{tab:Moon}
\end{center}
\end{table}

\section{Conclusions and Future Work}
In this paper, we implemented a two-dimensional implementation of the
Torquato-Jiao Adaptive Shrinking Cell scheme using a stochastic search
with simulated annealing to study the dense packing behavior of a
variety of well-known and nontrivial particle shapes.  We confirmed
the utility of the algorithm by reproducing the well-known densest
packings of regular pentagons and octagons.  Next, we applied the
algorithm to several nontrivial particle shapes to find their densest
packing behavior.  It is especially worth noting that the ASC scheme
correctly and accurately predicted subtle perturbations in the
two-particle packings of curved triangles that are not immediately
intuitively apparent.

\begin{figure*}[t!]
  \begin{centering}
    $\begin{array}{c@{\hspace{0.3in}}c@{\hspace{0.3in}}c}
      \includegraphics[width=0.25\textwidth]{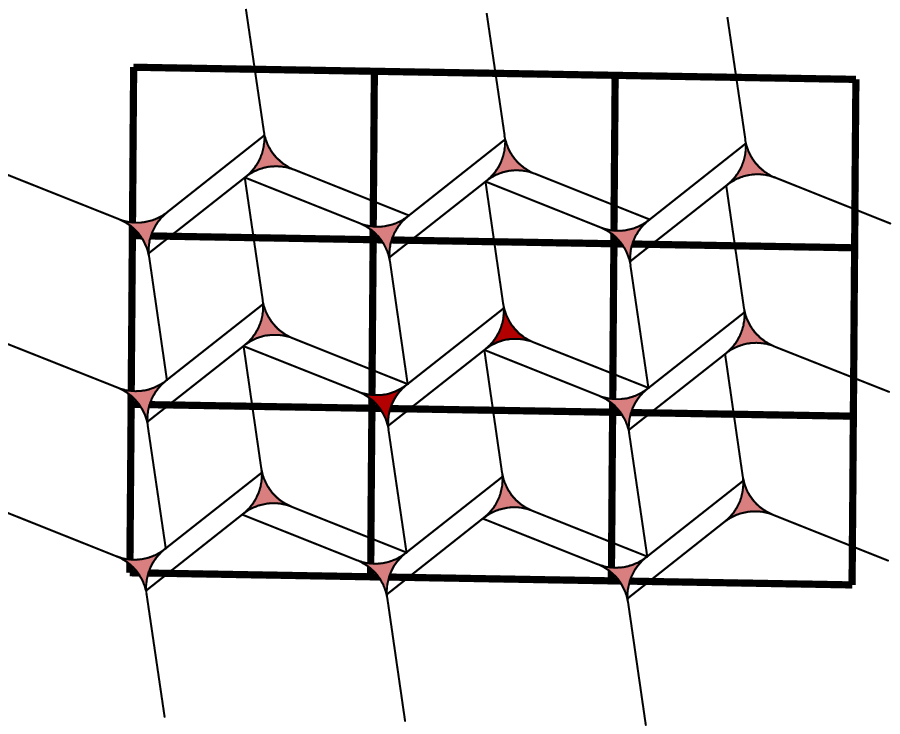}
      &
      \includegraphics[width=0.25\textwidth]{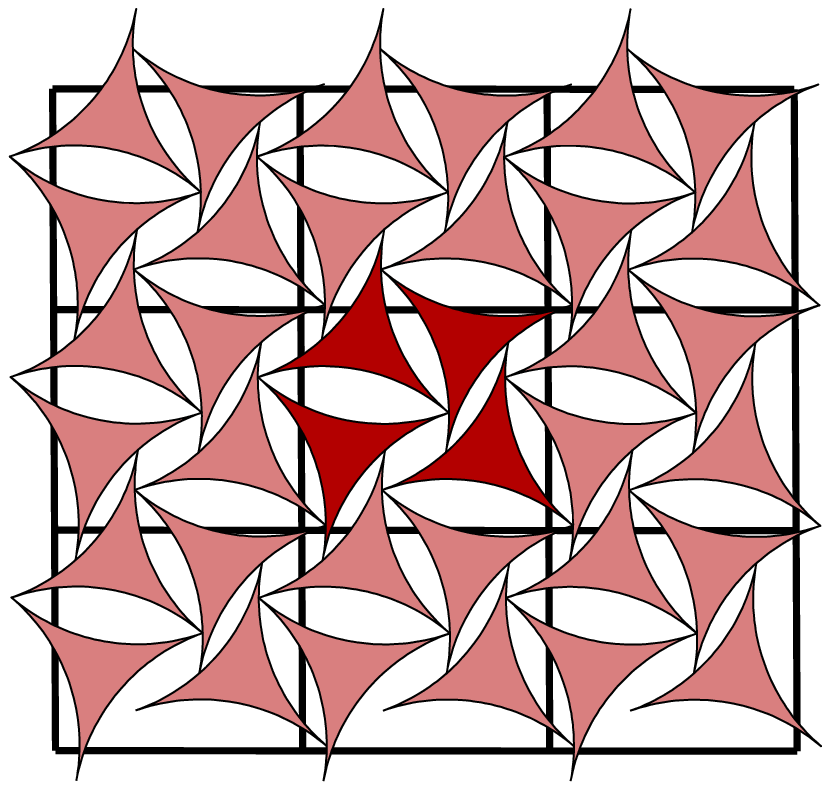}
      &
      \includegraphics[width=0.25\textwidth]{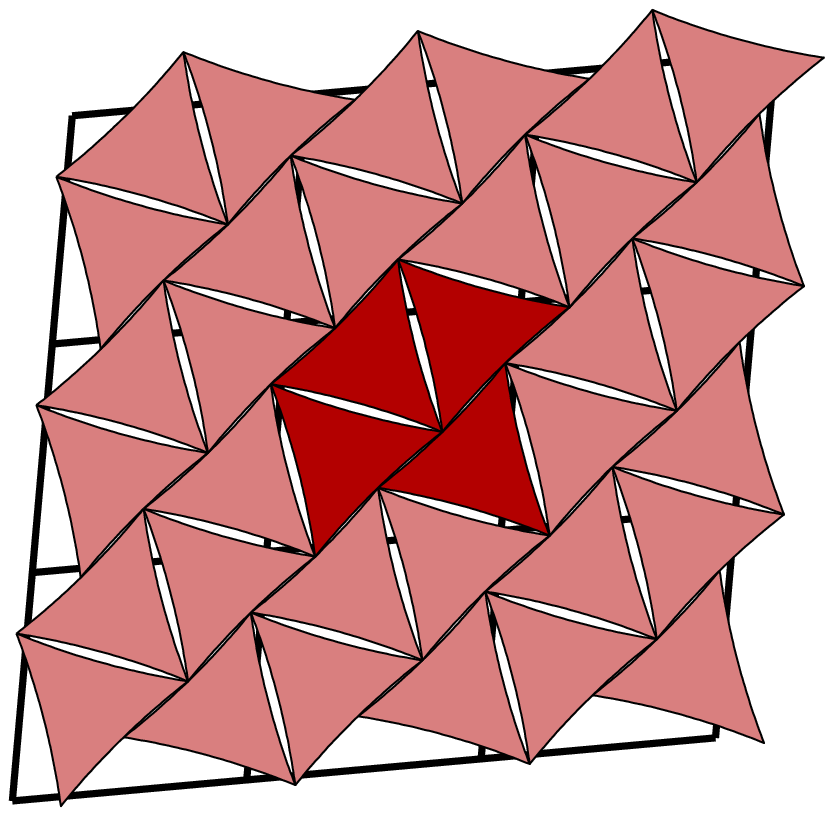}
      \\
      \mbox{(a)}
      &
      \mbox{(b)}
      &
      \mbox{(c)}
    \end{array}$
    \caption{(Color online) Computer-generated dense periodic packings
      of concave curved triangles for $k' = 1/20$ (a), $27/100$ (b),
      and $9/20$ (c).  The packing densities are $\phi = 0.019141$,
      $0.446702$, and $0.896363$, respectively. All of these packing
      suggest double-lattice structures.}
    \label{fig:ConcaveTriangles}
  \end{centering}
\end{figure*}

The packing characteristics of the curved triangle shape class are
unlike anything studied before because its optimal packing density is
{\it not} achieved using a double-lattice packing for sufficiently
high curvature.  In addition, the $B$-structure that achieves the
densest packings of certain moon-like shapes is a counterintuitive
finding for its lack of symmetry in its fundamental basis.  These
discoveries underscore the utility of the stochastic search implementation
of the ASC scheme because of its ability to accurately predict these
structures up to the small details inherent in them.

The packing structures discussed in this work offer insights towards
the organizing principles for three-dimensional particles in
Ref.\ \cite{OrganizingPrinciples}.  For example, the densest packings
of regular polygons, as two-dimensional analogs of convex polyhedra,
show similar behavior in that the densest packings of regular polygons
possessing central symmetry are given by their corresponding densest
lattice packings; and the densest packings of those lacking central
symmetry are nonetheless given by packings that possess a point of
inversion symmetry.  Furthermore, the densest packings of fat crosses
are given by their corresponding densest lattice packings, in much the
same way that the densest packings of three-dimensional,
centrally-symmetric polyhedra are conjectured to be given by their
corresponding densest lattice packings.  One final interesting remark
is that the $B$ structure of moon-like shapes possesses no points of
inversion.  This is in contrast to the proposition in three dimensions
that the densest packings of concave polyhedral particles are composed
of centrally symmetric compound units.  It seems intuitively true
that, by approximating the appropriate moon-like shape as a polygon
with sufficiently many edges, some structure similar to $B$ structure
will achieve the densest packing (which has no points of inversion).

It will be desirable in future work to determine the analytical
packing behavior of the concave instance of the curved triangle (for
which computer-generated results are provided in the Appendix) for the
sake of completeness.  More generally, the algorithm used in this work
may be adapted to generate both packings of hard particles in closed
containers.  By altering the shrinking behavior of the domain, it may
be possible to study both densest packings and disordered jammed
packings.  Furthermore, the algorithm may be readily adapted to study
packings of particles with a polydispersity in size and shape.
Finally, the stochastic search solution to the ASC scheme may readily
be adapted to investigate the dense packing behavior of concave solids
in higher dimensions.

\begin{figure}[h!]
  \begin{centering}
    \psfrag{k}{$k'$}
    \psfrag{rho}{$\phi$}
    \includegraphics[width=0.3\textwidth]{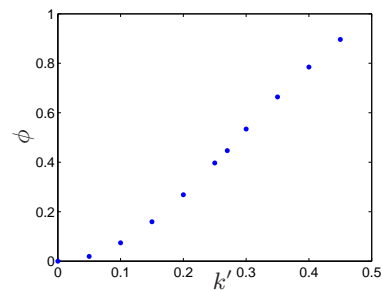}
    \caption{(Color online) Densities of computer generated packings
      of concave curved triangles for various values of $k'$.}
    \label{fig:ConcaveDensity}
  \end{centering}
\end{figure}

\section*{Acknowledgements}
This work was supported by the Materials Research Science and
Engineering Center (MRSEC) Program of the National Science Foundation
under Grant No. DMR-0820341 and by the Division of Mathematical
Sciences at the National Science Foundation under Award Number
DMS-1211087.

\section*{Appendix: Maximally dense packings of two-dimensional convex
  and concave noncircular particles}

Figure \ref{fig:ConcaveTriangles} shows a collection of
computer-generated dense periodic packings of {\it concave} curved
triangles with two- and four-particle bases.  A double-lattice
structure was observed for all particle shapes that we tried.  The
packing densities of the computer-generated cases are plotted in
Figure \ref{fig:ConcaveDensity} as a function of an alternative
parameter $k'$, defined as the ratio of the inradius of the particle
to its circumradius.  According to this alternative parameter, the $k'
= 0$ limit is equivalent to the $k = - \infty$ limit (Mercedes-Benz
limit), and $k' = 1/2$ describes an equilateral triangle ($k = 0$).


%

\end{document}